\documentclass[lettersize,journal]{IEEEtran}
\usepackage{amsmath,amsfonts}
\usepackage{amssymb}
\usepackage{algorithmic}
\usepackage{algorithm}
\usepackage{array}
\usepackage[caption=false,font=normalsize,labelfont=sf,textfont=sf]{subfig}
\usepackage{textcomp}
\usepackage{stfloats}
\usepackage{url}
\usepackage{threeparttable}
\usepackage{verbatim}
\usepackage{graphicx}
\usepackage{bbding}
\usepackage{orcidlink}
\usepackage{pifont}
\usepackage{wasysym}
\usepackage[]{cite}
\usepackage{enumitem}
\usepackage{titlesec}
\usepackage{tikz}
\usepackage{booktabs}
\usepackage{multirow} % 引入multirow宏包

\hypersetup{
	colorlinks=true,
	linkcolor=black,
	citecolor=black
}
\begin{document}

\title{Efficient and Secure Cross-Domain Data-Sharing for Resource-Constrained Internet of Things}
%AEAuth: Adaptive and Efficient Authentication for IoT in Cloud-Edge-Device Environments
\author{Kexian Liu$^{~\orcidlink{0000-0003-3975-5606}}$, Jianfeng Guan$^{~\orcidlink{0000-0002-4411-0741}}$,~\IEEEmembership{Member,~IEEE}, Xiaolong Hu, Jianli Liu, Hongke Zhang$^{~\orcidlink{0000-0001-8906-813X}}$,~\IEEEmembership{Fellow,~IEEE}
	\thanks{This work was supported by the National Key R\&D Program of China under Grant No. 2022YFB3102304 and in part by National Natural Science Foundation of China Grants(62394323, 62225105, 62001057) \textit{(Corresponding authors: Jianfeng Guan.)}
		
		Kexian Liu, Jianfeng Guan, Xiaolong Hu, Jing Zhang, and Jianli Liu are with the State Key Laboratory of Networking and Switching Technology, Beijing University of Posts and Telecommunications, Beijing 100876, China (e-mail: kxliu@bupt.edu.cn; jfguan@bupt.edu.cn; hxl814446051@bupt.edu.cn).
		
		Hongke Zhang is with the School of Electronic and Information Engineering, Beijing Jiaotong University, Beijing 100044, China (e-mail: hkzhang@bjtu.edu.cn).
	}% <-this % stops a space
}
\markboth{Journal of \LaTeX\ Class Files,~Vol.~14, No.~8, August~2021}%
{Shell \MakeLowercase{\textit{et al.}}: A Sample Article Using IEEEtran.cls for IEEE Journals}
\maketitle

\markboth{Journal of \LaTeX\ Class Files,~Vol.~14, No.~8, August~2021}%
{Shell \MakeLowercase{\textit{et al.}}: A Sample Article Using IEEEtran.cls for IEEE Journals}


\maketitle

\begin{abstract}
The growing complexity of Internet of Things (IoT) environments, particularly in cross-domain data sharing, presents significant security challenges. Existing data-sharing schemes often rely on computationally expensive cryptographic operations and centralized key management, limiting their effectiveness for resource-constrained devices. To address these issues, we propose an efficient, secure blockchain-based data-sharing scheme. First, our scheme adopts a distributed key generation method, which avoids single point of failure. This method also allows independent pseudonym generation and key updates, enhancing authentication flexibility while reducing computational overhead. Additionally, the scheme provides a complete data-sharing process, covering data uploading, storage, and sharing, while ensuring data traceability, integrity, and privacy. Security analysis shows that the proposed scheme is theoretically secure and resistant to various attacks, while performance evaluations demonstrate lower computational and communication overhead compared to existing solutions, making it both secure and efficient for IoT applications.
\end{abstract}


\section{Introduction}
\IEEEPARstart{T}{th} rapid advancement of networks has significantly benefited various domains. Simultaneously, these fields have developed increasingly complex requirements. Data exchange and collaboration across different domains are essential for achieving higher operational efficiency\cite{liu2023blockchain}\cite{yao2022blockchain}. For example, in the Industrial Internet of Things (IIoT), cross-domain data sharing, such as dimensions and specifications, enables precise production\cite{peter2023industrial,sun2023comprehensive}. Similarly, in the Medical Internet of Things (MIoT), doctors need to share real-time patient data, such as blood pressure and glucose levels, to make more accurate diagnoses\cite{mao2023locally,li2023efficient}. Likewise, in vehicular networks, cars must share location and incident data to ensure road safety\cite{zhong2023secure,hu2024security}. Fig. 1 shows a typical cloud-based data-sharing framework for the cross-domain IoT. Evidently, data sharing has become a cornerstone for the IoT's enhanced societal contribution. 

However, the data-sharing process is fraught with security challenges. For example, in 2019, attackers gained access to a family's WiFi password, enabling them to monitor the household and ultimately share private videos online without authorization. Likewise, in April 2024, SOCRadar’s security researchers discovered a data leak involving Microsoft's Azure cloud service, where sensitive information stored on public servers was exposed. The root cause was unauthorized access during sensitive data sharing, leading to malicious attacks and breaches. Thus, efficient and secure data-sharing methods are imperative to ensure the IoT's safe and reliable operation.

\begin{figure}[!t]
	\centering
	\includegraphics[width=0.5\textwidth]{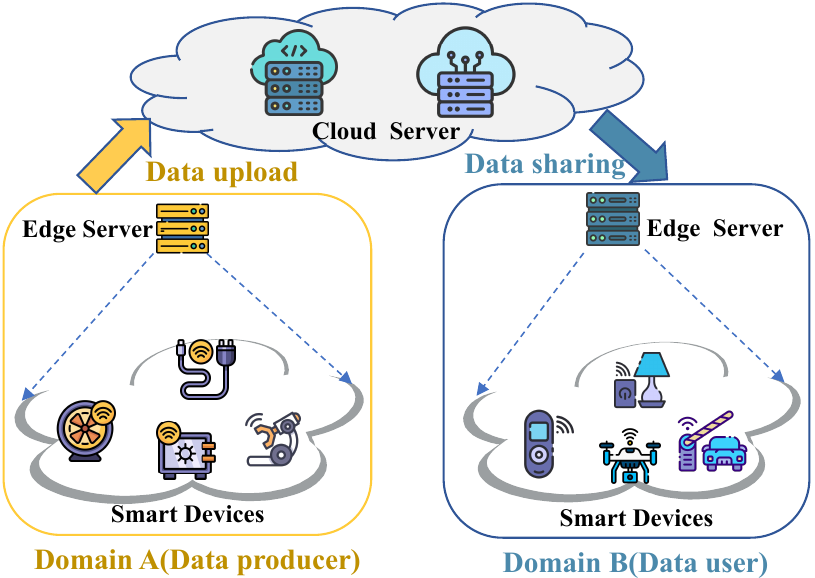}
	\caption{Cloud-based data-sharing framework in IoT.}
	\label{fig_1}
	\vspace{-1.6em}
\end{figure}

Several data-sharing schemes have been proposed for the IoT, including Attribute-Based Encryption (ABE) \cite{niu2023attribute,guo2023attribute}, searchable encryption\cite{liu2023key}, and Proxy Re-Encryption (PRE) schemes\cite{zhong2021broadcast,zhang2023identity}. Nevertheless, these methods have inherent drawbacks. Many rely on computationally expensive cryptographic operations, such as bilinear pairings, which impose a heavy computational burden on resource-constrained IoT devices. Additionally, these schemes often fail to fully address the auditing of data storage, access, and processing\cite{wang2023lightweight}. Moreover, key management typically depends on a single authority, introducing key security risks.\par 
Blockchain-based data-sharing approaches can mitigate key management issues and facilitate auditing. For instance, Liu \textit{et al.}\cite{liu2024dkgauth} and Cui \textit{et al.}\cite{cui2022efficient} proposed a blockchain-based cross-domain data exchange framework that alleviates key security concerns. However, their approach requires caching numerous pseudonyms generated by the Trusted Authority (TA) for smart devices (SDs), complicating the authentication process, reducing flexibility, and introducing efficiency challenges in pseudonym updates. Wang \textit{et al.}\cite{wang2023lightweight} proposed a blockchain-based data-sharing scheme for the cloud-edge-end architecture in the IIoT environment, which achieved data auditability. However, this solution also suffers from key and pseudonym update complexities, data traceability issues, and overly complex entity interactions. Later, Wang \textit{et al.}\cite{wang2024blockchain} enhanced this scheme to enable secure cross-domain data sharing in cloud-edge-end IIoT environments. Unfortunately, the updated approach relies on computationally expensive bilinear pairings, rendering it unsuitable for resource-constrained devices, and does not adequately address data uploading and traceability processes.

Thus, there is an urgent need for a secure and efficient data-sharing scheme in the IoT.

This research focuses on designing an efficient, secure, and comprehensive data-sharing scheme for resource-constrained IoT devices. The scheme aims to address the limitations of existing solutions and confronts the following challenges:

\begin{itemize}
	\item[$\bullet$] How to design a secure and flexible key and pseudonym management scheme that enhances smart device authentication while protecting device privacy?
	
	\item[$\bullet$] How to develop a complete, lightweight data-sharing protocol, suitable for resource-constrained devices, that ensures security throughout the processes of data uploading, storage, and sharing?
	
\end{itemize}\par

To address these challenges, we propose a blockchain-based distributed key management method that enhances the security and flexibility of the authentication process, and a complete, lightweight, blockchain-based data-sharing scheme, aided by edge computing, that ensures security throughout the entire data-sharing process.

The main contributions of this research are as follows:

\begin{itemize}
	\item[$\bullet$] We propose a blockchain-based cross-domain anonymous authentication method, which adopts a distributed key generation approach to achieve multi-domain key management. It enables SDs to generate their own pseudonyms for communication without relying on pre-generated ones from the TA. Additionally, SDs can independently update their pseudonyms and keys, reducing complex operations on the device side, making the scheme well-suited for resource-constrained SDs.
	
	\item[$\bullet$] We develop a blockchain-based and efficent data-sharing protocol that includes processes such as data uploading, storage, and sharing. The protocol binds the SD’s anonymous identity during data uploads, ensuring data traceability while preserving device privacy. Additionally, our scheme supports batch request verification, improving data-sharing efficiency. Leveraging the immutability of blockchain, we ensure data integrity and enhance security throughout the entire sharing process.
	
    \item[$\bullet$] We conducted a security analysis of our proposed scheme, and the results demonstrate that our scheme is theoretically secure and can resist certain attacks. Additionally, we analyzed the computational and communication overhead of our scheme. The results show that our scheme outperforms other solutions in terms of both security performance and efficiency.
	
\end{itemize}\par

The remainder of the paper is organized as follows. In Section II, we review the existing authentication and data sharing mechanisms in IoT environment. Section III provides preliminaries and background. Section IV provides system model of our scheme. In Section V, we elaborate on our scheme. Section VI conducts a security performance analysis of our scheme. In Section VII, we establish an experimental environment to evaluate the performance of the proposed scheme and compare it with other alternatives. Finally, Section VIII concludes the paper.

\section{Related work}

In this section, we discuss the challenges faced by authentication and data-sharing mechanisms in cross-domain IoT schemes.

\subsection{Authentication Mechanisms} 
In recent years, several public-private key-based authentication schemes have emerged in the IoT. These schemes can be broadly categorized into  public key infrastructure (PKI)-based, identity-based cryptography (IBC)-based, combined public key (CPK)-based, certicateless public key cryptography (CL-PKC)-based, and blockchain-based approaches.\par 
Since PKI was first introduced by Kohnfelder\cite{kohnfelder1978towards}, it has matured into a widely adopted authentication method across various fields\cite{koisser2022v,wang2020collaborative,chen2021xauth}. However, it still faces challenges, such as certificate authority (CA) failures, complex certificate management, and high authentication overhead, making it unsuitable for resource-constrained IoT environments. To address the complexity of PKI-based certificate management, IBC, CPK, and CL-PKC schemes have been proposed. Nonetheless, these approaches also present various limitations. In IBC and CPK systems\cite{hofheinz2024identity,zhang2022cpk}, key management is centralized by a single key generation center (KGC) or key management center (KMC), which creates a single point of failure, leading to severe key leakage incidents\cite{guan2021bsla,liu2024dkgauth}. On the other hand, in CL-PKC schemes\cite{fan2023anonymous}, the issue of public key replacement may arise\cite{pu2024generic}, resulting in identity spoofing risks.

To resolve the limitations inherent in these authentication architectures, blockchain-based authentication frameworks have been proposed. Shen \textit{et al.} \cite{shen2020blockchain} combined blockchain with IBC to enable cross-domain authentication in the IIoT. While this approach successfully established inter-domain trust and addressed issues related to certificate management, it still suffers from the problem of centralized key management by a single KGC, which can lead to single point of failure. Cui \textit{et al.} \cite{cui2022efficient} proposed a blockchain-based cross-domain authentication scheme for IIoT, where a single KGC generates numerous pseudonyms and keys for users or devices, and the blockchain proxies the consensus of pseudonym and public key. Although this approach mitigates the single point of failure issue to some extent, the key generation is still performed by a centralized entity.

The DKGAuth scheme\cite{liu2024dkgauth} takes this further by distributing key generation across multiple blockchain nodes and achieving cross-domain collaborative key management, effectively addressing the single point of failure in centralized key management. However, both Cui\cite{cui2022efficient}'s and DKGAuth schemes require devices to have numerous pseudonyms pre-generated by a certification authority and cached before authentication, which increases the device's burden and reduces the flexibility of the authentication scheme.

Some schemes have explored enabling devices to generate their own pseudonyms\cite{shen2020blockchain,guan2021bsla}, but these still face challenges in key management and computational overhead. Therefore, there remains a need for a flexible, secure, and efficient authentication solution in IoT environments.

\subsection{Data Sharing Mechanisms} 

In recent years, a series of research efforts have been proposed to achieve secure and efficient data sharing, including but not limited to ABE, searchable encryption, PRE, and blockchain-based technologies.

ABE-based data sharing schemes ensure not only data confidentiality but also fine-grained access control. For example, Guo \textit{et al.} \cite{guo2023attribute} proposed an ABE-based data sharing scheme for blockchain-enabled 6G Vehicular Ad Hoc Networks (VANETs), supporting policy hiding, data revocation, and cross-domain data sharing. Similarly, Ning \textit{et al.} \cite{ning2020dual} presented a cloud-based data storage and sharing scheme that uses ABE to authenticate Data Users (DUs). Searchable encryption is another critical technology in data sharing, enabling keyword search over encrypted data, thereby improving data owners' ability to selectively share encrypted information with users. Liu \textit{et al.} \cite{liu2023key} introduced a key-aggregate searchable encryption framework that supports conjunctive queries and offers two security models, enabling flexible data sharing and joint keyword search over encrypted data.

To combine the strengths of ABE and searchable encryption, researchers have designed hybrid schemes. For instance, Niu \textit{et al.} \cite{niu2023attribute} developed an efficient ABE-based searchable encryption scheme for edge computing, considering computational efficiency for both regular and lightweight users by outsourcing partial decryption operations as needed, making it more practical for real-world applications. Similarly, Zhang \textit{et al.} \cite{zhang2023blockchain} and Xiong \textit{et al.} \cite{xiong2023attribute} also integrated searchable encryption with ABE to achieve secure and efficient data sharing.

PRE technology enables ciphertext conversion without compromising data privacy. Ge \textit{et al.} \cite{ge2023attribute} proposed an ABE-based PRE with direct revocation (ABPRE-DR) mechanism for encrypted data sharing, allowing cloud servers to directly revoke user keys from the original shared data set. Pei \textit{et al.} \cite{pei2024proxy} presented a secure data-sharing solution based on PRE technology, enhancing security in the IoMT environment. Zhang \textit{et al.} \cite{zhang2023identity} and Zhong \textit{et al.} \cite{zhong2021broadcast} proposed broadcast PRE schemes for VANETs, achieving flexible and efficient data sharing in vehicular networks.

Although the above studies ensure data security through various techniques, many of them involve computationally expensive cryptographic operations, such as bilinear pairing, which impose a heavy computational burden on resource-constrained IoT devices. Additionally, these schemes do not fully achieve the auditability of data storage, access, and processing.

To address the traceability and auditability of data storage and access, some researchers have introduced blockchain into IoT environments\cite{miyachi2021hocbs,guan2021bsla}. Chen \textit{et al.} \cite{chen2020blockchain} proposed a lightweight PRE scheme that leverages blockchain and equality test technology to ensure secure data sharing. However, in this scheme, smart devices directly store data on the blockchain, which increases the storage burden and limits its applicability to IoT environments with small data volumes. Furthermore, this scheme does not focus on protecting the anonymity of smart devices. Lu \textit{et al.} \cite{lu2021blockchain} proposed a blockchain-based data storage and sharing scheme that combines group signatures and PRE technology to achieve secure storage and access authentication, ensuring both device anonymity and data source trustworthiness. However, group signatures involve numerous computationally intensive operations, placing a significant computational load on the blockchain-based data-sharing platform.

Wang \textit{et al.} \cite{wang2023lightweight} introduced a blockchain-based data-sharing scheme for the Industrial Internet, achieving auditability for data storage and access. However, the scheme suffers from key and pseudonym update challenges, as well as overly complex entity interactions in data sharing. Later, They \cite{wang2023lightweight} proposed blockchain-based secure cross-domain data sharing for IIoT, which enables secure cross-domain data sharing. Unfortunately, this scheme uses complex bilinear pairing operations, making it unsuitable for resource-constrained devices and omitting critical details regarding the data upload process.

In summary, current data-sharing schemes face several limitations and challenges when applied to IoT environments. Therefore, designing a comprehensive, lightweight, auditable, and traceable data-sharing scheme is of paramount importance.

\section{PRELIMINARIES AND BACKGROUND}
\subsection{Blockchain}
Blockchain technology and its ecosystem originated from Bitcoin. With the concentrated research and adoption of blockchain technology in various industries including finance, supply chain, culture and entertainment, and the IoT, its product model is also constantly evolving. This technology's unique characteristics, including distribution and tamper-proof features, enable its application in a broader spectrum of distributed collaborations. During the experiment, we modify the existing platform to function as an on-chain execution platform. Simultaneously, we translate the smart contract language into the compiler of the on-chain platform.
\subsubsection*{\bf What we have done:}
\begin{itemize}
	\item[$\bullet$] We design a well-structured distributed ledger, which includes creating a powerful consensus mechanism to store data, ensuring that important information is kept in the system, preventing unauthorized changes to devices' private content, and recording all access operations for audit.
\end{itemize}
\begin{itemize}
	\item[$\bullet$] We enable interactive implementation of the blockchain platform by empowering platform nodes to deploy contracts and execute management functions, including querying, modifying, and deleting data on the chain.
\end{itemize}

\subsection{Elliptic Curve Cryptosystem(ECC)}
The elliptic curve cryptosystem uses rational points on the elliptic curve to establish the computational complexity of elliptic discrete logarithms in the Abel additive group. Its main advantage is that it can use shorter keys than other cryptographic systems. While providing enhanced security.
The equation of the elliptic curve is given by $y^2 = x^2 + ax + b\ mod\ p$, where $a$ and $b$ are adjustable parameters and $p$ is a large prime number. These parameters define various generator curves. Specifically,based on this base curve, the additive cyclic group of order $q$ with the generator $P$ has the following characteristics.

\begin{itemize}
	\item[$\bullet$]Elliptic curve discrete logarithm problem (ECDLP): Given a base point $P$, a positive integer $s$ is randomly chosen within its operating field. In this context, we have $Q = sP$, where $Q$ serves as the public key and $s$ as the private key. While it is relatively straightforward to obtain $Q$ when knowing both $s$ and $P$, deducing $s$ from $Q$ and $P$ poses a challenging mathematical problem inherent to elliptic curve cryptography.
\end{itemize}

\begin{itemize}
	\item[$\bullet$] Elliptic curve computational Diffie-Hellman problem (ECCDHP):\ If an attacker knows $P$, $Q_1 = s_1P$, and $Q_2 = s_2P$, computing $s_1s_2P$ poses a mathematical challenge.
\end{itemize}

\begin{itemize}
	\item[$\bullet$] ECC Composite Theorem:
	Within ECC,  it has been proven that given multiple pairs of public and private keys, their respective sums can form a new pair of public and private keys. For instance, the sum of private keys can be expressed as: $(s_1 + s_2 + ... + s_n)\ mod\ q = s$, which corresponds to the sum of public keys: $Q_1 + Q_2 + ... + Q_n = Q$ . Thus, $s$ and $Q$ precisely form a new pair of public and private keys. 
\end{itemize}

\section{SYSTEM MODEL}

\subsection{Network Model and Assumptions}

Figure 2 illustrates the proposed system data sharing system model, which primarily comprises smart devices (SDs), edge servers (ESs), cloud servers (CSs), and blockchain (BC). The functionalities and system assumptions of these entities are as follows:

{\bf{\textit{1) Cloud servers (CSs)}}}: CSs provide various network services for devices, equipped with ample storage and computational resources. In this study, CSs are used to store the encrypted data shared among SDs.

{\bf{\textit{2) Edge Servers (ESs)}}}: ESs can communicate with other entities such as SDs and CSs, but each ES can only communicate with SDs within its range. It is assumed that ESs have strong computational capabilities but weaker storage capacities compared to CSs. ESs can collect and upload data from SDs within their communication range. Additionally, ESs across different domains jointly maintain a distributed ledger (BC) (refer to DKGAuth\cite{liu2024dkgauth}). ESs in each domain also act as registration and key management centers, collaboratively managing the network’s keys with ESs from other domains. It is assumed that ESs are trusted and not compromised.

{\bf{\textit{3) Smart Devices (SDs)}}}: SDs possess good wireless communication capabilities but have limited computational power and storage space. They can upload data and act as data generators. SDs from different domains can also download data from CSs, making them both data requesters and users.

{\bf{\textit{4) Blockchain (BC)}}}: The BC is a distributed ledger maintained collectively by ESs from various domains. It primarily stores the hash values of the original data and the indices of the data stored on CSs. Data can be shared with all valid ESs.

\subsection{System Workflow}
\begin{figure}[!t]
	\centering
	\includegraphics[width=0.5\textwidth]{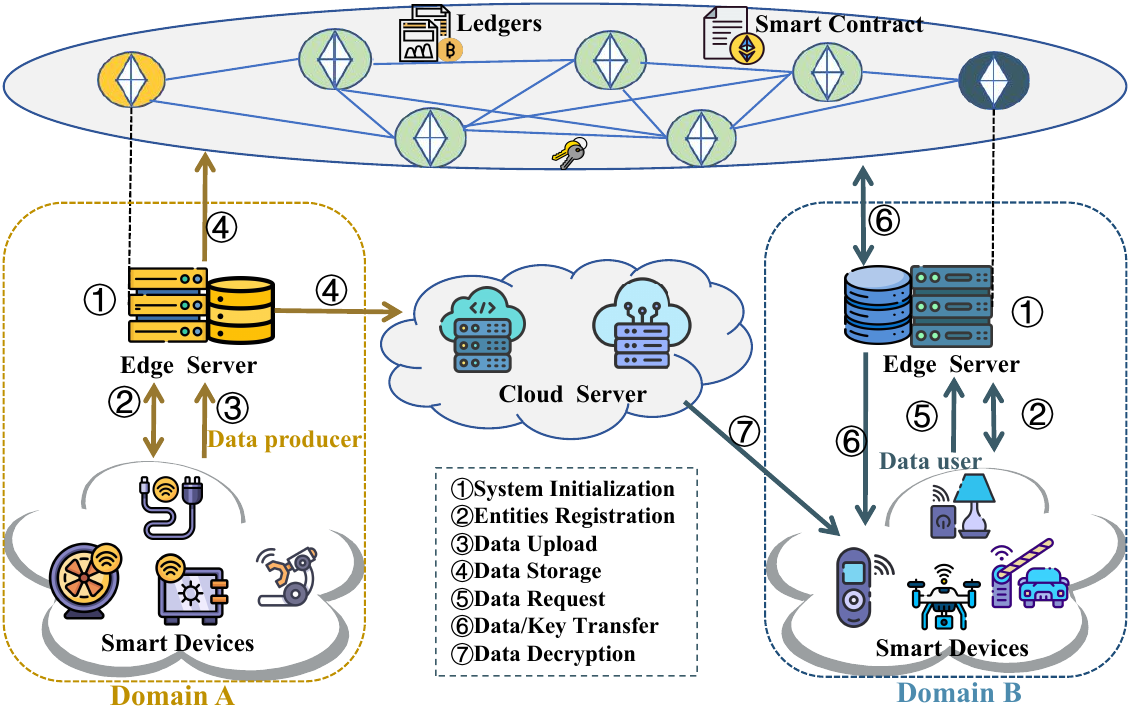}
	\caption{System model.}
	\label{fig_1}
	\vspace{-1.6em}
\end{figure}

Our proposed solution consists of eight steps:

{\bf{\textcircled{1}System Initialization.}} In this phase, the system initializes its ECC public key parameters and Hash functions. These parameters are publicly announced for use by entities within the network.

{\bf{\textcircled{2}Entities Registration.}}
Entities within the network register with the registration center (ES) of their respective domain. The registration center generates the necessary materials for authentication and encryption for the registering entities.

{\bf{\textcircled{3}Data Upload.}}
In this phase, SDs generate data to be shared and upload it to the ES of their domain. During this process, SDs sign and encrypt the data they are uploading.

{\bf{\textcircled{4}Data Storage.}}
Upon receiving the uploaded data, the ES authenticates the data and then processes it for storage, storing the data on both the CS and the BC.

{\bf{\textcircled{5}Data Request.}}
In this phase, data users send their data requests along with a signed authorization.

{\bf{\textcircled{6}Data/Key Transfer.}}
In this phase, upon receiving a data request, the ES verifies the requester’s identity. It then encrypts the index and key from the BC and returns it to the data user.

{\bf{\textcircled{7} Data Decryption.}}
Upon receiving the encrypted index and key from the ES, the SD decrypts them. It then downloads the original encrypted data from the CS and verifies the data’s legitimacy.

{\bf{\textcircled{8}Revoking/Reviewing Abnormal Devices.}}
If the data user detects illegal data, the anonymous identity associated with the illegal data is reported back to the ES of the domain. The ES traces the data back to the uploader for review.

\subsection{Security and Privacy Requirements}

\begin{itemize}
	\item[$\bullet$] {\bf{\textit{Correctness}}}: The integrity of the original data is ensured, preventing any tampering by attackers.
\end{itemize}

\begin{itemize}
	\item[$\bullet$] {\bf{\textit{Key Security}}}: The security of the cryptographic keys must be maintained to prevent unauthorized access and ensure that the keys cannot be compromised by attackers. The keys should always be kept secure, and the system must avoid single point of failure in key management centers.
\end{itemize}

\begin{itemize}
	\item[$\bullet$] {\bf{\textit{Data Traceability}}}:The data uploaded by each SD is bound to its identity, enabling the ES to trace the data back to the corresponding SD. 
\end{itemize}

\begin{itemize}
	\item[$\bullet$] {\bf{\textit{Confidentiality}}}: The confidentiality of the original data must be protected against attacks from malicious actors.
\end{itemize}

\begin{itemize}
	\item[$\bullet$] {\bf{\textit{Anonymity}}}: To prevent the privacy of SDs from being compromised, SDs use anonymous identities for uploading and downloading data. Their real identities should not be known to entities outside of the ES.
\end{itemize}

\begin{itemize}
	\item[$\bullet$] {\bf{\textit{Unlinkability}}}: Entities can randomly generate pseudonyms each time they upload or request data, accompanied by different random values. Therefore, the communication is generally unlinkable.
\end{itemize}

\begin{itemize}
	\item[$\bullet$] {\bf{\textit{Resilience Against Common Attacks}}}: The proposed scheme should be able to resist common types of attacks, such as replay attacks, impersonation attacks and single point of failure, ensuring the security of the entire network.
\end{itemize}

\section{OUR PROPOSED SCHEME}

\renewcommand\arraystretch{1.4}
\begin{table}[h]
	\caption{Summary of abbreviations and notation}
	\centering  
	\begin{tabular}{p{2.4cm}p{5.6cm}}
		\hline
		\textbf{Notation} & \textbf{Description}\\ 
		\hline
		$TA$ & Trusted authority \\
		$SD_i$ & i-th smart device \\
		$ES_j,CS_k$ &  j-th ES and  k-th CS, respectively \\
		$DID_i$ & Real identities of $D_i$\\
		$PID_i$ & Pseudonym of $SD_i$ \\
		${SK_i,PK_i}$ & Private-public key sequences of $SD_i$ \\
		$psk_{i},ppk_{i}$  & Partial private-public key of $SD_i$\\
		$sk_{i},pk_{i}$  & Private-public key of $SD_i$\\
		$SV_i$ & Secret values of $SD_i$\\
		$M_i,C_i$ & Message upload and its ciphertext from $SD_i$\\
		$key_i$ & Key for encrypting $M_i$ and decrypting $C_i$. \\
		$index_i$ & Index where $C_i$ is stored. \\
		$s,P_{pub}$ & System private key and publice key.\\
		$\theta_i,\sigma_j$ & Signature of $SD_i$,$SD_j$.\\
		$\alpha_i,\beta_j,RV_i,V_i,HR_3$ & Hash values\\
		$r_i,r_j,R_1,R_2,R_3$ & Temporary Values\\
		$T_i,T_j,T_k$ & Timestamp\\
		$H_i(i=1,2,...,9)$ & Collision-resistant hash function\\
		$\oplus$ & XOR operation\\
		\hline
	\end{tabular}
\end{table}
In this section, we describe our proposed scheme in detail. As introduced in Section VI, the System Workflow, our scheme comprises the following stages: System Initialization, Entities Registration, Data Upload, Data Storage, Data Request, Data/Key Transfer, Data Decryption, and Revoking/Reviewing Abnormal Devices.

Table I lists the main symbols used and their corresponding definitions throughout these eight stages. Our scheme offers several advantages: 1)Reduced Communication Overhead: Devices can update keys without interacting with the TA, significantly reducing the communication overhead between devices and the TA. 2)Enhanced Privacy and Traceability: Data uploaded by devices is encrypted and uploaded anonymously, enhancing both the privacy of the devices and the data. At the same time, the scheme supports data traceability. 3)Efficient Data Sharing: The uploaded data is encrypted by the device only once, eliminating the need for complex proxy re-encryption. This allows multiple data users to decrypt and use the data, thereby increasing the efficiency of data sharing.

\subsection{System Initialization}
\begin{itemize}
	\item[1)] Each ES jointly maintains a distributed ledger, also known as a blockchain, and initiates a consensus protocol to ensure its security and consistency. In this context, the blockchain maintained by the ESs can be referred to as the TA.

	\item[2)] The TA begins by selecting a random prime $p$ to establish a finite field $\mathbb{F}_p^*$. Then, it randomly chooses two elliptic curve parameters $a$ and $b$,  where $a,b \in \mathbb{F}_p^*$. An elliptic curve $E$ is then generated, represented as $y^2 = x^3 + ax + b$ mod $p$. The TA also selects another prime $q$ and a point $P$ to generate the group $\mathbb{G}$.
	
	\item[3)] The TA randomly selects a secret key $s$, where $s \in \mathbb{Z}_q^*$ and computes the public key $P_{pub} = sP$. It is important to note that $s$ is locally kept secret by the ESs in each domain and is not accessible to any entities other than the ESs.

	\item[4)] A series of secure hash functions $H_1: \{0, 1\}^* \times \mathbb{Z}_q^* \to \mathbb{Z}_q^*$, $H_2: \mathbb{G} \to \{0, 1\}^*$, $H_3: \{0, 1\}^* \times \{0, 1\}^* \times \{0, 1\}^*\to \{0, 1\}^*$, $H_4: \mathbb{Z}_q^*\times\mathbb{G}\times\{0, 1\}^* \to \{0, 1\}^* $, $H_5:\mathbb{G}\times \{0, 1\}^*\times\{0, 1\}^*\times\{0, 1\}^*\times\{0, 1\}^* \to \mathbb{Z}_q^* $, $H_6: \mathbb{G}\times \{0, 1\}^*\times\{0, 1\}^*\times\{0, 1\}^*\times\{0, 1\}^*\times\{0, 1\}^* \to \{0, 1\}^*$, $H_7: \mathbb{G}\times\{0, 1\}^* \to \{0, 1\}^*$, $H_8: \{0, 1\}^*\times\{0, 1\}^*\times\{0, 1\}^* \to \mathbb{Z}_q^*$, $H_9: \{0, 1\}^*\times\{0, 1\}^* \to \{0, 1\}^*$ are chosen.
	\item[5)] The public parameters $\{P_{pub},H_i(i=1,2,...,9)\}$ will be published by the TA.
	
\end{itemize}

\subsection{Entities Registration}

In this phase, the SD provides its registration information to the TA through a secure channel for identity registration.
First, the device selects $ES_i$ of its domain as the registration center. 
\begin{itemize}

  \item[1)] $SD_i$ sends its decentralized identity $DID_i$ to the $ES_i$.

  \item[2)] Upon receiving the $DID_i$, the $ES_i$ first verifies its legitimacy, such as checking for duplicate registrations. If the $DID_i$ is invalid, an error message is returned to $SD_i$. If valid, $ES_i$ binds the $DID_i$ with publicly available random ECC parameters on the blockchain, achieving on-chain consensus. Each blockchain node (ES) generates partial public and private key sequences based on the $DID_i$ and its associated public random ECC parameters: $SK_i= \{psk_1, psk_2, ...\}$, $PK_i = \{ppk_1, ppk_2, ...\}$ (For further details, please refer to our previous work, DKGAuth\cite{liu2024dkgauth}). Additionally, $ES_i$ computes $SV_i = H_1(DID_i, s)$. Finally, $ES_i$ returns the partial private key sequence $SK_i$ and the secret value $SV_i$ to $SD_i$, while the partial public key sequence $PK_i$ is stored on the blockchain.\par
  It is important to note that to save on-chain storage resources, $SD_i$'s $PK_i$ can be stored on a cloud server, with only their indices and verification information stored on the blockchain. To protect $SD_i$ privacy, only ESs have the permission to access and query the blockchain, while ordinary SDs do not have viewing privileges.

  \item[3)] Upon receiving $\{SK_i, SV_i\}$, $SD_i$ securely stores them locally.
	
\end{itemize}

\subsection{Data Upload}

After completing registration, $SD_i$ operates for a period and generates some data that needs to be shared. To facilitate data sharing, the $SD_i$ needs to upload the generated data. $SD_i$ will first randomly generate a pseudonym and key. Then, it will encrypt and sign the generated message $M_i$, and subsequently send the signature along with the relevant information to $ES_i$ in its domain. The upload steps are as follows:

\begin{itemize}
\item[1)] $SD_i$ randomly selects a value $r_i$ from $\mathbb{Z}_q^*$ and computes $PID_{i,1} = r_iP$ and $R_2 = r_iP_{pub}$. Next, $SD_i$ calculates $PID_{i,2} = DID_i \oplus H_2(R_2)$ and the key $sk_i$ using the key generation algorithm $sk_i = DKGAuth(PID_{i,2},SK_i)$\cite{liu2024dkgauth}. The pseudonym $PID_i = \{PID_{i,1}, PID_{i,2}\}$.

\item[2)] Assuming the data $SD_i$ needs to upload is $M_i$, it computes $RV_i = H_3(M_i, PID_i, T_i)$, where $T_i$ is the timestamp when $M_i$ is generated.
It then calculates the encryption $key = H_4(SV_i, R_2, T_i)$ and encrypts the data $M_i$ to obtain $C_i = ENC_{key}(M_i, PID_i, T_i)$. 

\item[3)] Then, it computes $\alpha_i = H_5(PID_i, C_i, RV_i, T_i)$ and the signature $\theta_i = r_i + \alpha_i sk_i$.

\item[4)] $SD_i$ sends the message $\{M_{Type}, PID_i, \theta_i, C_i, RV_i, T_i\}$ to the $ES_i$, where $M_{Type}$ is the service type of $M_i$.

\end{itemize}

\subsection{Data Storage}
In this phase, after collecting the data sent by $SD_i$, $ES_i)$ first checks if the message has expired. Then, it verifies the $SD_i$'s signature to confirm the legitimacy of its identity. Upon successful verification, the ES stores the data in both the CS and the BC. Additionally, we support batch authentication.

\begin{itemize}
\item[1)] After receiving the message $\{M_{Type}, PID_i, \theta_i, C_i, RV_i, T_i\}$, $ES_i$ first checks if $T_i$ has expired. If it has not expired, it computes $R_2' = s PID_{i,1}$ and determines the $SD_i$'s real identity $DID_i = PID_{i,2} \oplus H_2(R2')$. Using the $DID_i$, it locates the corresponding patial public key sequence $PK_i$ and calculates $pk_i = DKGAuth(PID_{i,2},PK_i)$\cite{liu2024dkgauth}. Then, it computes $\alpha_i' = H_5(PID_i, C_i, RV_i, T_i)$ and calculates whether(1) $\theta_i P = PID_{i,1} + \alpha_i' pk_i$ holds. The process of calculation is shown below. 
\begin{equation}
	\begin{split}
	\theta_i P &= (r_i + h(PID_i, C_i, RV_i, T_i)sk_i)P\\
	         &= r_iP + \alpha_i sk_i\\
	         &= PID_{i,1} + \alpha_i pk_i
	\end{split}      
\end{equation}

\item[2)] If the signature is valid, it stores $C_i$ in the CS and obtains its index value $index_i$. It then computes $V_i = H_6(PID_i, C_i, RV_i, index_i, T_i)$ and $pindex_i = H_7(R_2',T_i) \oplus index_i$.

\item[3)] According to the service type $M_{Type}$, it stores $\{M_{Type}, PID_i, V_i, pindex_i, T_i\}$ on the BC.
\end{itemize}

{\bf{\textit{Batch Verification}}}: Additionally, we support batch message verification to reduce verification latency. Suppose $n$ SDs need to upload data simultaneously, resulting in multiple upload requests such as $ Msg_1 = \{M_{Type1}, PID_1, $$\theta_1, C_1, RV_1, T_1\}$, $Msg_2 = \{M_{Type2}, PID_2,$$ \theta_2, C_2, RV_2, T_2\}$, $...Msg_n = \{M_{Typen}, PID_n, \theta_n, $$C_n, RV_n, T_n\}$, where the timestamps are valid. $ES_i$ begins batch authentication after obtaining the signature $\theta$ and the corresponding public keys of the SDs. 

\begin{itemize}
	\item[1)] To prevent batch verification attacks, $ES_i$ randomly selects a vector $v = \{v_1, v_2, ..., v_i, ..., v_n\}$, where $v_i \in [1, 2^\zeta]$, and $\zeta$ is a small random integer\cite{he2015efficient}.
	\item[2)] Then $ES_i$ verifies signatures by (2).
	
	\begin{equation}
		\begin{split}
			(\sum_{i=1}^{n} v_i \theta_i)P = (\sum_{i=1}^{n} v_i PID_{i,1}) + (\sum_{i=1}^{n} v_i \alpha_i pk_i)
		\end{split}      
	\end{equation}
The process of proving (2) as follows:
	\begin{equation}
		\begin{split}
			(\sum_{i=1}^{n} v_i \theta_i)P 
			&= (\sum_{i=1}^{n} v_i (r_i + \alpha_i sk_i))P\\
			&= (\sum_{i=1}^{n} v_i r_iP + (\sum_{i=1}^{n} v_i\alpha_i sk_i)P\\
			&= (\sum_{i=1}^{n} v_i PID_{i,1}) + (\sum_{i=1}^{n} v_i \alpha_i pk_i)
		\end{split}      
	\end{equation}
\end{itemize}

\subsection{Data Request}

When a SD in another domain wants to request data of a certain service type, it sends a data request to the ES of its domain. The SD first generates a pseudonym and a key and signs the request information. This process is similar to data uploading, as detailed below:

\begin{itemize}
	
\item[1)] $SD_j$ randomly selects $r_j$ from $\mathbb{Z}_p^*$ and computes $PID_{j,1} = r_jP$ and $HR_3 = H_2(r_jP_{pub})$. Then, $SD_j$ calculates $PID_{j,2} = DID_j \oplus HR_3$ and generates the key $sk_j$ using the DKGAuth algorithm\cite{liu2024dkgauth}, $sk_j = DKGAuth(PID_{j,2}, SK_j)$. The pseudonym $PID_j = \{PID_{j,1}, PID_{j,2}\}$.
\item[2)] It computes $\beta_j = H_8(M_{Type}, PID_j, T_j)$ and the signature $\delta_j = r_j + \beta_j sk_j$.
\item[3)] $SD_j$ sends the message $\{M_{Type}, PID_j,\delta_j, T_j\}$ to the ES of its domain.

\end{itemize}

\subsection{Data/Key Transfer}
In this phase, after receiving the $SD_j$'s data request, $ES_j$ first verifies the signature to confirm the $SD_j$'s identity. Then, based on the subscription information, it retrieves the relevant data information from the BC. Finally, it encrypts this information and sends it back to the $SD_j$. The detailed steps are as follows:

\begin{itemize}
\item[1)] Upon receiving the message $\{M_{Type}, PID_j,\delta_j, T_j\}$, $ES_j$ first checks if $T_j$ has expired. If it has not expired, it computes $R_3' = s PID_{j,1} $ and $HR_3' = H_2(R_3')$. It then determines the $SD_j$'s true identity $DID_j = PID_j \oplus HR_3$. Using $DID_j$, it locates the corresponding public key factor sequence $PK_j$ and calculates $pk_j = DKGAuth(PID_{j,2},PK_j)$\cite{liu2024dkgauth}.\par
It is important to note that once the ES verifies the true identity of the data requester, it will check if the requester has subscribed to the service by confirming their access permissions. If the verification fails, the request will be rejected; if it succeeds, the process will proceed.
\item[2)]Then, $ES_j$ computes $\beta_j' = H_8(M_{Type}, PID_j, t_j)$ and calculates whether $\theta_j P = PID_{j,1} + \beta_j' pk_j$ holds. 
\item[3)] If the signature is valid, $ES_j$ queries the chain for the relevant information $\{M_{Type}, PID_i, V_i, pindex_i, T_i\}$. It then computes $R_2' = s PID_{i,1}$ and $DID_i = PID_{i,2} \oplus H_2(R_2')$, and the encryption key $key' = H_4(H_1(DID_i, s), R_2', T_i)$. Next, $ES_j$ calculates $k = key'\oplus H_9(HR_3',T_j)$ and $pindex' = pindex \oplus H_7(R_2',T_i) \oplus H_9(HR_3',T_j)$.
\item[4)] Finally, $ES_j$ returns the message $\{pindex', V_i, k,T_k\}$ to $SD_j$.
\end{itemize}

It should be noted that $ES_j$ can also handle batch authentication for requests from multiple SDs, similar to Equation 2 mentioned above. This process is not elaborated here.

\subsection{Data Decryption}

In this phase, $SD_j$ decrypts and verifies the data received from the $ES_j$. First, it retrieves the index of the original encrypted data stored in the CS, then obtains the original encrypted data based on the index, and finally decrypts and verifies the data. The detailed steps are as follows:
\begin{itemize}
	\item[1)] Upon receiving $\{pindex', V_i, k,T_k\}$, $SD_j$ first checks $T_k$ to determine whether the message has expired. If it has expired, the message is discarded. If it has not expired, $SD_j$ calculates $index’ = pindex’ \oplus H_9(HR_3,t_j)$. Using $index’$, it retrieves the ciphertext $C_i'$ from the CS.
	\item[2)] Then, $SD_j$ calculates the encryption key $key'' = k \oplus H_9(HR_3,t_j)$ and decrypts the ciphertext $C_i$ to obtain the plaintext $\{M_i', PID_i', T_i'\} = DEC_{key''}(C_i')$.
	\item[3)] Finally, $SD_j$ calculates $RV’ = h(M_i', PID_i', T_i')$ and $V_i' = h(PID_i', C_i', RV_i', index_i', T_i')$. If $V_i' = V_i$, the key and metadata are verified as correct. If not, it indicates tampering, and the incident is reported to the administrator.
\end{itemize}

\subsection{Revoking/Reviewing Abnormal Devices}

In this phase, if the data user discovers an anomaly and suspects the data to be problematic, meaning that the original data uploaded by the producer is illegal, he user will send the $PID_i$, obtained by decrypting the data, to the ES of its domain.

The ES then calculates $DID_i = PID_{i,2} \oplus h(s PID_{i,1})$ to determine the true identity of the data producer, and subsequently conducts further review and processing.

\section{SECURITY PROOF AND ANALYSIS}

This section primarily analyzes and proves the security of our proposed anonymous authentication scheme and data-sharing protocol. Our scheme mainly consists of two parts: (1) the security of SDs signature algorithm, and (2) the confidentiality of the encryption keys and data indexes. Therefore, we will separately prove the security of these two parts.

\subsection{Security Proof}

{\bf{{1. Security of the SDs Signature Algorithm}}}

We will demonstrate that, under the assumption that the Elliptic Curve Discrete Logarithm Problem (ECDLP) cannot be solved in polynomial time, our scheme can securely prevent forgery under an adaptive chosen-message attack in the random oracle model. The security model of the signature scheme is defined by a game played between an adversary $\mathcal{A}$ and a challenger $\mathcal{C}$, where $\mathcal{A}$ can make the following queries during the game:

\begin{itemize}
	
	\item[1)]\textit{Setup - Oracle}: $\mathcal{C}$ selects the security parameters and the system private key, and outputs the system parameters and the system public key to $\mathcal{A}$.
	
	\item[2)]\textit{$H_2$ - Oracle}: When $\mathcal{A}$ calls this oracle, $\mathcal{C}$ returns a randomly chosen string $str \in \{0, 1\}^*$ to $\mathcal{A}$ and inserts the tuple $\{ m, str \}$ into $L_1$.
	
	\item[3)]\textit{$H_5$ - Oracle}: When $\mathcal{A}$ calls this oracle, $\mathcal{C}$ returns a randomly chosen number $x \in \mathbb{Z}_q^*$ to $\mathcal{A}$ and inserts the tuple $\{ m, x \}$ into $L_2$.
	
	\item[4)]\textit{Sign - Oracle}: When $\mathcal{A}$ calls this oracle with a message $M_i$, $\mathcal{C}$ returns the signature $\{\theta_i, M_i, PID_i, T_i \}$ to $\mathcal{A}$.

\end{itemize}

$\mathcal{A}$ wins the game if it outputs a signature $\theta*$ and $m*$ that satisfy the verification equation (1) and $m*$ has not been previously signed by the Sign. Let $\mathcal{A}$ be an adversary that, within a time limit $t'$, can execute a forgery attack against our scheme with a probability $\Phi$ under an adaptive chosen-message attack. Theorem 1 of our scheme is shown as follows.

{\textit{Theorem 1 }}\cite{pointcheval2000security}: Let $q_{H_2}$, $q_{H_5}$, and $q_{sign}$ denote the number of queries that adversary $\mathcal{A}$ can make to the oracles $H_2$, $H_5$ and $Sign$, respectively, within time $t$. If adversary $\mathcal{A}$ can forge a signature scheme with a non-negligible advantage $\Phi$, then there exists a challenger $\mathcal{C}$ that can solve the ECDLP within time $t'$, expected to be less than 120686$q_{h1} q_{h2} t'/\Phi$, provided that $\Phi \ge 10 (q_{sign} + 1)(q_{H_2} + q_{H_5} + q_{sign}) / 2^n$.

{\bf{\textit{Proof}}}: Assume that $\mathcal{A}$, as the forger, can forge a valid signature tuple. At the same time, challenger $\mathcal{C}$ executes the scheme using a random instance $\{sk_i, pk_i\}$ and provides $sk_i \in \mathbb{Z}_q^*$. We will  show that challenger $\mathcal{C}$ can solve the ECDLP with a non-negligible probability by using $\mathcal{A}$ as a subroutine.

\begin{itemize}

	\item[1)]\textit{Setup}: Challenger $\mathcal{C}$ runs the setup algorithm using the security parameter. $\mathcal{C}$ generates an elliptic curve of order $q$, selects a generator point $P$, and randomly chooses $sk$ and $pk$ as the system's private and public keys, respectively. Subsequently, $\mathcal{C}$ generates the parameters $\{P, q, a, b, sk, pk, s, P_{pub} H_2, H_5\}$ and sends them to $\mathcal{A}$. Note that $\mathcal{C}$ maintains the Query-Oracle lists $L_1$, $L_2$ to track $\mathcal{A}$'s queries to the hash oracles $H_2$ and $H_5$. Initially, these lists are empty.
	
	\item[2)]\textit{$H_2$ Hash Query}: The format of the $H_2 - Oracle$ list $L_1$ is $\{r_xP_{pub}, \tau_{H_2}\}$. If $\mathcal{A}$ queries $H_2$ with the tuple $\{r_xP_{pub}\}$, $\mathcal{C}$ checks whether $L_1$ contains $\{r_xP_{pub}, \tau_{H_2}\}$. If it does, $\mathcal{C}$ returns the value $\tau_{H_2}$ to $\mathcal{A}$. Otherwise, $\mathcal{C}$ randomly selects $\tau_{H_2}\in \mathbb{Z}_q^*$ and adds the new tuple $\{r_xP_{pub}, \tau_{H_2}\}$ to the hash list $L_1$. Then, $\mathcal{C}$ returns the value $\tau_{H_2} = H_2(r_xP_{pub})$ to $\mathcal{A}$.
	
	\item[3)]\textit{$H_5$ Hash Query}: The format of the $H_5 - Oracle$ list $L_2$ is $\{PID_i, C_i, RV_i, T_i, \tau_{H_5}\}$. If $\mathcal{A}$ queries $H_5$ with the tuple $\{PID_i, C_i, RV_i, T_i\}$, $\mathcal{C}$ checks whether $L_2$ contains $\{PID_i, C_i, RV_i, T_i, \tau_{H_5}\}$. If it does, $\mathcal{C}$ returns the value $\tau_{H_5}$ to $\mathcal{A}$. Otherwise, $\mathcal{C}$ randomly selects $\tau_{H_5}\in \mathbb{Z}_q^*$ and adds the new tuple $\{PID_i, C_i, RV_i, T_i, \tau_{H_5}\}$ to the hash list $L_2$. Then, $\mathcal{C}$ returns the value $\tau_{H_5} = H_5(PID_i, C_i, RV_i, T_i)$ to $\mathcal{A}$.
	
	\item[4)]\textit{Sign Query}: Upon receiving a sign query with message $M_i$ from $\mathcal{A}$, $\mathcal{C}$ randomly selects $\theta_i$ and $\alpha_i \in \mathbb{Z}_q^* $ and computes $PID_{i,1} = \theta_i P - \alpha pk$. Then $\mathcal{C}$ inserts the tuple $\{PID_i, C_i, RV_i, T_i, \tau_{h2}\}$ into $L_2$. Subsequently, $\mathcal{C}$ returns $\{PID_i, C_i, RV_i, T_i, \theta_i\}$ to $\mathcal{A}$. According to the game's rules, all responses to sign queries are valid signatures, which can be verified using the equation: 
		\begin{equation}
		\begin{split}
			\theta_i P &= PID_{i,1} + \alpha_i pk\\
			&= (\theta_i P - \alpha_i pk) + \alpha_i pk\\
			&= \theta_i P
		\end{split}      
	\end{equation}
\end{itemize}
Thus, $\mathcal{A}$ can generate valid signed messages with non-negligible probability.
Using the Forking Lemma\cite{pointcheval2000security}, by sending the same elements to $\mathcal{A}$ again, $\mathcal{C}$ can generate two valid signed messages $\{PID_i, C_i, RV_i, T_i, \theta_i\}$ and $\{PID_i, C_i, RV_i, T_i, \theta_i^*\}$.
\begin{equation}
	\theta_iP = (r_i + \alpha_i sk_i)P.  
\end{equation}
\begin{equation}
	\theta_i^*P = (r_i + \alpha_i^* sk_i)P.  
\end{equation}

From equations (5) and (6), we get the following equation:

\begin{equation}
	(\theta_i - \theta_i^*) P = (\alpha_i - \alpha_i^*)sk_iP
\end{equation}

Consequently, $\mathcal{C}$ can output $(\theta_i - \theta_i^*)(\alpha_i - \alpha_i^*)^{-1}$ as the solution to the ECDLP within expected time less than 120686$q_{H_2} q_{H_5} t'/\Phi$, provided that $\Phi \ge 10 (q_{sign} + 1)(q_{H_2} + q_{H_5} + q_{sign}) / 2^n$. This contradicts Definition 1. Therefore, the proposed scheme is secure against adaptive chosen-message attacks in the random oracle model.

{\bf{{2.Confidentiality of the Encryption Key and Data Index}}}

The confidentiality of the original data, encryption key, and data index relies on the Elliptic curve computational Diffie-Hellman problem (ECCDHP) and the collision resistance of the hash function.

In the proof, we use $m = (key, index)$ to represent the plaintext $m$. The execution process of the ciphertext indistinguishability game is as follows:

\begin{itemize}
	
	\item[1)]\textit{Setup}: $\mathcal{C}$ selects the security parameters and the system private key, and outputs the system parameters and the system public key to $\mathcal{A}$.
	
	\item[2)]\textit{Ciphertext Oracle O}: $\mathcal{A}$ adaptive launches some queries to $\mathcal{C}$ as the following oracle. Given a plaintext message $m$, the oracle generates the encryption key and ciphertext $\{k, pindex\}$.
	
	\item[3)]\textit{Challenge}: $\mathcal{A}$ selects two plaintext messages (m0, m1) and sends them to $\mathcal{C}$. $\mathcal{C}$ chooses a random bit $c \in \{0, 1\}$, then generates the corresponding ciphertext $\{k, pindex\}$. Finally, $\mathcal{C}$ sends $\{k, pindex\}$ to $\mathcal{A}$.
	
	\item[4)]\textit{Oracle O query}: $\mathcal{A}$ can initiate queries to $O$. Note that $\mathcal{A}$ cannot submit $m_0$ or $m_1$ to the $oracle O$.
	
	\item[5)]\textit{Guess}: $\mathcal{A}$ outputs a bit $c' \in \{0, 1\}$. If $c' = c$, $\mathcal{A}$ wins the game. $\mathcal{A}$'s advantage in the game is defined as 
	
	\begin{equation}
	Adv_{\mathcal{A}}^{IND}(\lambda) = \arrowvert Pr[c' = c] - \frac{1}{2} \arrowvert \nonumber
	\end{equation}
	
\end{itemize}\par
{\textit{Theorem 2}}: In the random oracle model, if the ECCDHP problem is intractable and the hash function is collision-resistant, the proposed scheme achieves ciphertext indistinguishability.

{\bf{\textit{Proof}}}: We prove the indistinguishability of the ciphertext through a series of related games. In the final game, we show that the ciphertext and the plaintext message are independent.

{\textit{$Game_0$}}: This game is the same as the ciphertext indistinguishability game. Challenger $\mathcal{C}$ first generates system parameters ${P, h_1}$ and two pairs of public/secret keys $\{s, P_{pub}\}$ and $\{sk_i, pk_i\}$. Then, $\mathcal{C}$ publishes the public data $\{pk_i, P_{pub}\}$ and keeps the secret keys $\{s, sk_i\}$. In response to a ciphertext oracle query, for $m = \{key, index\}$, $\mathcal{C}$ computes $PID_{i,1} = r_iP$, $R_2 = r_1P_{pub}$, and $key = H_4(SV_i, R_2, T_i)$. Then, $\mathcal{C}$ generates the ciphertext $pindex = H_7(R_2',T_i) \oplus index$ and $k = key \oplus H_9(HR_3',t_j)$. This process simulates $\mathcal{A}$'s ability to obtain the corresponding ciphertext. After a series of such queries, $\mathcal{A}$ guesses the challenge ciphertext and ultimately wins the game with an advantage of 

\begin{equation}
	Adv_{\mathcal{A}}^{Game_0}(\lambda) = Adv_{\mathcal{A}}^{IND}(\lambda) \nonumber
\end{equation}

{\textit{$Game_1$}}: The game is the same as $Game_0$ except it uses a secure hash function $H_2$. The $H_2$-Oracle list $L_1$ is formatted as $\{r_xP_{pub}, \tau_{H_2}\}$. When $\mathcal{A}$ queries $H_2$ with $\{r_xP_{pub}\}$, $\mathcal{C}$ checks if $\{r_xP_{pub}, \tau_{H_2}\}$ exists in $L_1$. If it does, $\mathcal{C}$ returns $\tau_{H_2}$ to $\mathcal{A}$. If not, $\mathcal{C}$ randomly selects $\tau_{H_2}\in \{0,1\}^*$, adds the tuple $\{r_xP{pub}, \tau_{H_2}\}$ to $L_1$, and returns $\tau_{H_2} = H_2(r_xP_{pub})$ to $\mathcal{A}$. But the hash function $H_2$ is secure, we have:

\begin{equation}
	Adv_{\mathcal{A}}^{Game_1}(\lambda) = Adv_{\mathcal{A}}^{Game_0}(\lambda) \nonumber
\end{equation}

{\textit{$Game_2$}}: This game is similar to $Game_1$, but instead of computing $pindex = H_2(r_xP_{pub}) \oplus index$ and $k = key \oplus H_2(r_xP_{pub})$, the challenger $\mathcal{A}$ chooses a random string $S \in \{0, 1\}*$ and then computes $pindex' = S \oplus index$ and $k' = key \oplus S$ to obtain the corresponding ciphertext $\{pindex, k\}$. In $Game 2$, given an ECCDHP instance $\{P, r_xP_{pub}\}$, adversary $\mathcal{A}$ does not know $\{s, r_x\}$ and cannot solve the ECCDHP with non-negligible probability, and thus cannot distinguish between $k'$ and $k$, as well as between $pindex'$ and $pindex$. Therefore, $\mathcal{A}$'s advantage in this game is:
\begin{equation}
	Adv_{\mathcal{A}}^{Game_2}(\lambda) - Adv_{\mathcal{A}}^{Game_1}(\lambda) \le Adv_{\mathcal{A}}^{ECCDHP}(\lambda)\nonumber
\end{equation}

{\textit{$Game_3$}}: This game is similar to $Game_2$, but instead of computing $pindex = H_2(r_xP_{pub}) \oplus index$ and $k = key \oplus H_2(r_xP_{pub})$, the challenger $\mathcal{C}$ randomly selects $pindex',k' \in \{0, 1\}^*$. Since $S$ is random, from $\mathcal{A}$'s perspective, $pindex'$ and $pindex$, $k'$ and $k$ are indistinguishable. Hence, $\mathcal{A}$'s advantage in this game is:

\begin{equation}
	Adv_{\mathcal{A}}^{Game_2}(\lambda) = Adv_{\mathcal{A}}^{Game_3}(\lambda) \nonumber
\end{equation}

From $Game_3$, we see that the ciphertext is chosen randomly and is independent of the plaintext. Therefore, the advantage in the third game is zero:
\begin{equation}
	Adv_{\mathcal{A}}^{Game_3}(\lambda) = \arrowvert\frac{1}{2}-\frac{1}{2} \arrowvert = 0 \nonumber
\end{equation}
Thus, we can obtain 

\begin{equation}
	Adv_{\mathcal{A}}^{IND}(\lambda) \le Adv_{\mathcal{A}}^{ECCDHP}(\lambda) \nonumber
\end{equation}

According to Section III, we know that the $Adv_{\mathcal{A}}^{ECCDHP}(\lambda)$ is non-negligible. Therefore, in the random oracle model, the key and index in our proposed scheme are secure and confidential.

\subsection{Informal Security Analysis}

In this subsection, we provide informal analysis to demonstrate the security of our scheme.

\begin{itemize}
	
	\item[1)]\textit{Confidentiality}: The original data is encrypted locally by the data producer and remains in an encrypted state throughout its subsequent transmission. Unauthorized SDs and ESs cannot access it. Only when it reaches the data user will it be decrypted using the key. Thus, in our scheme, the original data remains confidential throughout the entire process.
	
	\item[2)]\textit{Key Security}: SD registration and public key information are maintained on a blockchain jointly managed by multiple ESs, achieving coordinated multi-domain key management and preventing single point of failure in the key management center.
	
	\item[3)]\textit{Data Traceability}: In our scheme, SDs bind pseudonyms with the data and upload it in encrypted form, such as $C_i=ENC_{key}(M_i,PID_i,T_i)$. When a data user finds incorrect data, they can report the pseudonym to the ES. The TA (ES) can then obtain the true identity $DID_i = PID_{i,2} \oplus H_2(sPID_{i,1})$ for auditing the actual SD.

	\item[4)]\textit{Perfect Forward and Backward Secrecy}: In our scheme, the key is computed as $key_i = H_4(H_1(DID_i,s), H_2(r_iP_{pub}), T_i)$. Even if an attacker obtains the key at a certain moment, they cannot derive the system key $s$. Therefore, the attacker cannot obtain the key for the next or previous moments, ensuring perfect forward and backward secrecy.

	\item[5)]\textit{Conditional Privacy Protection}: The SD's anonymous identity $PID_i$ is locally computed as $PID_{i,1} =r_iP$ and $PID_{i,2} = DID_i \oplus H_2(r_iP_{pub})$. Only the TA (ES) and the device itself can know its true identity. This allows the TA (ES) to trace the identity of malicious users while preserving SD anonymity, thus achieving conditional privacy protection.
	
	\item[6)]\textit{Anonymity}: In our scheme, SDs generate and use random pseudonyms when uploading and requesting data. The pseudonym $PID_i$ is composed of $PID_{i,1} =r_iP$ and $PID_{i,2} = DID_i \oplus H_2(r_iP_{pub})$, where $r_i$ is a random number. Based on the ECCDHP problem, attackers cannot derive the SD's true identity from the $PID_i$ without knowing the system secret key $s$.
	
	\item[7)]\textit{Unlinkability}: Our scheme uses random numbers and timestamps, such as $r_i$, $r_j$, $T_i$, $T_j$ and $T_k$ to ensure that messages transmitted over the network are different. Additionally, since the PIDs are randomly generated, attackers cannot determine if two different messages originate from the same sender.
	
	\item[8)]\textit{Resistance to Impersonation Attacks}: In our scheme, there are two types of impersonation attacks:
	
	\textit{SDs Impersonation Attack}: An attacker intending to impersonate $SD_i$ would need to obtain the genuine $SD_i$'s key to compute the correct signature $\theta_i = r_i + \aleph_i sk_i$. $SD_i$'s key is locally and secretly stored, making it inaccessible to the attacker. Additionally, when the ES verifies the signature and decrypts the true identity to obtain the corresponding public key information, the SDs communicates using pseudonyms, preventing the attacker from obtaining the true SDs information. Hence, impersonating SDs is impractical.
	
	\textit{ESs Impersonation Attack}: This attack occurs during the data request phase. When a data requester sends a signed message $\{M_{Type}, PID_j,\delta_j, T_j\}$ to the ES, the genuine ES verifies the signature. $\mathcal{A}$ fake ES may try to forge the information $\{M_{Type}, PID_i, V_i, pindex_i, T_i\}$ stored on the blockchain but cannot obtain the secret system key $s$ and therefore cannot generate a legitimate encryption $key_i = H_4(H_1(DID_i,s), H_2(sPID_{i,1})$ and $index_i = pindex_i \oplus H_2(sPID_{i,1})$. 
	Even if it sends forged $key'$ and $index'$ to the SD, it cannot obtain $r_jP_{pub}$ or $sPID_{j,1}$ to compute correct $k = key \oplus H_2(sPID_{j,1})$ and $pindex=index \oplus H_2(sPID_{j,1})$. When $SD_i$ receives the data, it uses $key = k \oplus H_2(r_jP_{pub})$ and $index=pindex \oplus H_2(r_jP_{pub})$ for decryption, and the forged $key$ and $index$ will fail verification, making ES impersonation impossible.
	
    \item[9)]\textit{Resistance to Replay Attacks}: We assume that an attacker $\mathcal{A}$ can monitor communications between entities. Although the attacker can intercept the message, the message contains timestamps and highly random numbers with short lifespans, such as $r_i$, $r_j$, $T_i$, $T_j$ and $T_k$. Hence, the proposed scheme offers certain defenses against replay attacks.
    
	\item[10)]\textit{Resistance to Modification Attacks}: In our scheme, both data uploaders and users sign the messages they send, such as $\{M_{Type}, PID_i, \theta_i, C_i, RV_i, T_i\}$, where $\theta_i = r_i + \alpha_i sk_i$ and $\alpha_i = H_5(PID_i, C_i, RV_i, T_i)$, verifying the message's integrity. Additionally, during the process of ES returning data-related information to the data user, when the ES returns $\{pindex', V_i, k\}$, where $pindex' = pindex \oplus H_2(s PID_{i,1}) \oplus H_2(s PID_{j,1})$, $V_i = H_6(PID_i, C_i, RV_i, index_i, T_i)$, and $k = key\oplus H_2(s PID_{j,1})$ the SD performs an integrity check with $RV’ = H_3(M_i, PID_i, T_i)$ and $V_i' = H_6(PID_i', C_i', RV_i', index_i', T_i')$. Therefore, our scheme can resist modification attacks during the data-sharing process.

\end{itemize}\par

\section{PERFORMANCE ANALYSIS}

In this section, we will perform a performance evaluation and analysis of our scheme including computational costs and communication overhead. We will also compare the performance of our scheme with other approaches.

\subsection{Experimental Settings}
In terms of environment configuration, We chose the FISCO-BCOS blockchain\cite{bl} system as the simulation platform, utilizing PBFT as the consensus protocol. To simulate a lightweight scenario, ES were deployed on a PC with 16GB RAM, running Ubuntu 16.04 as the operating system, and equipped with a 11th Intel Core i5-11400H CPU @2.7GHz(12 cores). To simulate IoT devices, we used individual virtual machine configured with a single-core CPU and 2GB of memory, which is comparable to the computational capabilities of most smart devices. The virtual machines were built using Workstation Pro 17.
In terms of algorithm implementation, we used MiRACL library\cite{MI}. The hash function used was SHA-256, and the elliptic curve $y^2 = x^3 + ax $$+ b (mod q)$ was employed, where $q$ is 256 bits prime number. And the keys based ECC is 256 bits. The symmetric bilinear pairs $e$ : $\mathbb{G}$ × $\mathbb{G}$ → $\mathbb{G}_T$ that we achieve 80-bit security level.
We discuss comparisons of computation and communication overheads in the authentication and key agreement phases of the proposed scheme and other existing related schemes, such as those by Zhong \textit{et al.} \cite{zhong2021broadcast},   Wang \textit{et al.} \cite{wang2023lightweight}, Mao \textit{et al.} \cite{mao2023locally}, Hu \textit{et al.} \cite{hu2024security} and Wang \textit{et al.} \cite{wang2024blockchain}.

\subsection{Computation Cost Analysis}

\begin{table}[h]
	\caption{Execution time of basic operations(ms)}
	\centering
	\begin{tabular}{p{0.9cm}p{3.7cm}p{1.1cm}p{1.3cm}}  % 这里设置了每一列的宽度
		\hline
		Operation & Description & Time(Edge) & Time(Device)\\ 
		\hline
		$T_{bp}$ & Bilinear pairing operation & 20.345 & 61.934\\
		$T_a$ & ECC point addition & 0.035 & 0.091\\
		$xT_a$ & $x$ times ECC point additions & 4.275 & 11.648\\
		$T_m$ & ECC point multiplication & 9.234 & 28.263\\
		$T_{sm}$ & Multiplication with small factor & 0.039 & 0.745\\
		$T_{e}$ & Exponentiation operation in $\mathbb{G}_T$ & 6.257 & 30.315\\
		$T_{mtp}$ & MapToPoint operation & 0.091 & 0.675\\
		$T_{gtmul}$ & Multiplication operation in $\mathbb{G}_T$ & 0.087 & 0.472\\
		$T_h$ & SHA-256 hash function & 0.001 & 0.004\\
		\hline
	\end{tabular}
\end{table}

\begin{table*}[h]
	\centering
	\caption{Your table caption here}
	\begin{tabular}{|>{\centering\arraybackslash}m{1cm}|>{\centering\arraybackslash}m{1cm}|>{\centering\arraybackslash}m{1.4cm}|>{\centering\arraybackslash}m{1.8cm}|>{\centering\arraybackslash}m{1.4cm}|>{\centering\arraybackslash}m{1.4cm}|>{\centering\arraybackslash}m{1.8cm}|>{\centering\arraybackslash}m{1.4cm}|>{\centering\arraybackslash}m{1.4cm}|>{\centering\arraybackslash}m{1.5cm}|}
		\hline
		\multirow{2}{*}{Scheme} & \multirow{1}{*}{$SD_i$} & \multicolumn{3}{|c|}{$ES_i$} & \multicolumn{3}{|c|}{$ES_j$} & \multicolumn{2}{|c|}{$SD_j$} \\
		\cline{2-10}
		& Upload Data & Verify\par Signature & Batch\par Verification & Collect/Store Data & Verify\par Signature & Batch\par Verification & Re-encrypt & Data\par Request & Decrypt\par Data \\
		\hline
		Zhong \textit{et al.} \cite{zhong2021broadcast} & - & - & - & - & $3T_{bp}+2T_m+T_a+T_{gtmul}+3T_h$ & $3T_{bp}+2nT_m+2nT_{sm}+(4n-3)T_{a}+3nT_h+4T_{gtmul}$ & $2T_{bp}+4T_m+2T_e+T_a+3T_{gtmul}+2T_h$ & $6T_m+2T_a+4T_h$ &  $2T_{bp}+T_{mtp}+2T_{gtmul}$\\
		\hline
		Wang \textit{et al.} \cite{wang2023lightweight} & $5T_m+5T_h$ & $4T_m+2T_h+2T_a$ & $nT_{sm} + (n+2)T_m+(2n+2)T_a+2nT_h$ & $5T_m+5T_h+2T_a$ & $2T_m+2T_h$ & - & $2T_m$ & $T_m+T_h$ & $3T_m+4T_h+T_a$ \\
		\hline
		Mao \textit{et al.} \cite{mao2023locally} & $T_{bp}+2T_h+T_m$ & $2T_{bp}+2T_m+T_a$ & $2T_{bp} + nT_a +(2n+1)T_m$ & $T_{bp}+T_{mtp}+T_h+T_e$ & - & - & - & - & - \\
		\hline
		Hu \textit{et al.} \cite{hu2024security} & $3T_m+T_h$ & $T_m+T_a+2T_{bp}$ & $2T_{bp}+nT_h+nT_m+(2n+1)T_a$ & $2T_{bp}+5T_h+9T_m+4T_a+T_e$ & $T_{bp}+2T_m+2T_a+T_h$ & - & - & - & - \\
		\hline
		Wang \textit{et al.} \cite{wang2024blockchain} & - & - & - & $T_{bp}+5T_m+2T_a+2T_{mtp}+T_e+4T_h$ & $4T_m+2T_a+T_h$ & $(2n+2)T_m+(2n+1)T_a+nT_h+nT_{sm}$ & $2T_{bp}+4T_m+T_a+2T_{mtp}+2T_h$ & $3T_m+T_a+T_h$ & $T_{bp}+T_m+T_{mtp}+2T_e+4T_h+T_{gtmul}$ \\
		\hline
		Ours & $2T_m+4T_h$ & $3T_m+2T_h+(x+1)T_a$ & $(n+1)T_m + nT_{sm}+(2n+xn+1)T_a+nT_h$ & $T_h$ & $3T_m+2T_h+(x+1)T_a$ & $(n+1)T_m + nT_{sm}+(2n+xn+1)T_a+nT_h$ & $T_m+4T_h$ & $2T_m+2T_h$ & $4T_h$ \\
		\hline
	\end{tabular}
	
\end{table*}
In this subsection, we will conduct a computational cost analysis of our scheme and compare it with other schemes. We analyze the relatively time-consuming cryptographic operations in the proposed solution, including Bilinear pairing operation, ECC point addition, $x$ times ECC point additions, ECC scalar multiplication, Multiplication with small factor, Exponentiation operation in $\mathbb{G}_T$, MapToPoint operation, Multiplication operation in $\mathbb{G}_T$ and SHA-256 hash are denoted by $T_{bp}$/$T_{a}$/$xT_{a}$/$T_{m}$/$T_{sm}/T_{e}/T_{mtp}/T_{gtmul}/T_{h}$. It should be noted that $xT_a$ refers to the time required for performing $x$ times ECC point additions of different elements, such as $PK_i + PK_j + PK_k + \ldots + PK_{i+x-1}$. Referring to the DKGAuth algorithm\cite{liu2024dkgauth}, the value of $x$ should range between 1 and 256. We randomly selected 1000 values of $x$ and calculated the average time for $xT_a$.
The other simulation experiment was repeated 100 times, and take the the average as the result. Table II shows the execution time of some basic operations used in this paper.\par

For security purposes, we assume that each time a device signs, it generates a new pseudonym and random numbers.

In Zhong \textit{et al.} \cite{zhong2021broadcast}’s scheme, the entire process includes signing, signature verification, encryption, re-encryption, and decryption. The signing incurs a cost of 6$T_m$ + 4$T_h$ + 2$T_a \approx 169.776$ ms, verification costs 3$T_{bp}$ + 2$T_m$ + 3$T_h$ + $T_a$ + $T_{gtmul} \approx 79.628$ ms, encryption costs $T_m$ + $T_h$ + $T_{gtmul} \approx 9.322$ ms, re-encryption costs 2$T_{bp}$ + 3$T_m$ + 2$T_e$ + $T_h$ + $T_a$ + 2$T_{gtmul} \approx 81.116$ ms, and decryption costs 2$T_{bp}$ + $T_{mtp}$ + 2$T_{gtmul} \approx 125.487$ ms.

In Wang \textit{et al.} \cite{wang2023lightweight}’s scheme, the process involves signing, signature verification, data packaging and storage, re-encryption, and decryption. The signing incurs a cost of 2$T_h$ + 3$T_m \approx 84.797$ ms, signature verification costs 2$T_h$ + 4$T_m$ + 2$T_a \approx 37.008$ ms, data packaging costs 2$T_m$ + 3$T_h \approx 18.471$ ms, storage costs 2$T_h$ + 3$T_m$ + 2$T_a \approx 27.774$ ms, data user signing costs $T_m$ + $T_h \approx 28.267$ ms, verification costs 2$T_m$ + 2$T_h \approx 18.470$ ms, re-encryption costs 2$T_m \approx 18.468$ ms, and decryption costs 4$T_h$ + 3$T_m$ + $T_a \approx 84.896$ ms.

In Mao \textit{et al.} \cite{mao2023locally}’s scheme, the process includes the signature computation by the data producer, signature verification by the data receiver, and decryption operations. The signing incurs a cost of $T_{bp}$ + 2$T_h$ + $T_m \approx 90.205$ ms, signature verification costs 2$T_{bp}$ + 2$T_m$ + $T_a \approx 59.193$ ms, and decryption costs $T_{bp}$ + $T_{mtp}$ + $T_e$ + $T_h \approx 26.694$ ms.

In Hu \textit{et al.} \cite{hu2024security}’s scheme, the process includes data generation and uploading, data verification, data analysis, and data storage and sharing. The data generation and uploading incur a cost of 3$T_m$ + $T_h \approx 84.793$ ms, data verification costs 2$T_{bp}$ + $T_a$ + $T_m \approx 49.959$ ms, data analysis costs 2$T_{bp}$ + 9$T_m$ + 4$T_a$ + 5$T_h$ + $T_e \approx 130.198$ ms, and data sharing costs $T_{bp}$ + 2$T_m$ + 2$T_a$ + $T_h \approx 38.884$ ms.

In Wang \textit{et al.} \cite{wang2024blockchain}’s scheme, the process involves data encryption and storage, signing, signature verification, key transfer, and decryption. Data encryption and storage incur a cost of $T_{bp}$ + 4$T_h$ + 5$T_m$ + 2$T_a$ + 2$T_{mtp}$ + $T_e \approx 73.028$ ms, signing incurs a cost of $T_h$ + 3$T_m$ + $T_a \approx 84.884$ ms, signature verification costs $T_h$ + 4$T_m$ + 2$T_a \approx 37.007$ ms, key transfer costs 2$T_{bp}$ + 2$T_h$ + 4$T_m$ + $T_a$ + 2$T_{mtp} \approx 77.845$ ms, and decryption costs $T_{bp}$ + $T_{mtp}$ + $T_m$ + 2$T_e$ + 4$T_h$ + $T_{gtmul} \approx 151.99$ ms.

In our proposed scheme, the process includes data uploading (including signing), ES data storage (including signature verification), data requests (including signing), data/key transformation (including signature verification), and data decryption. The data upload incurs a cost of 4$T_h$ + 2$T_m \approx 56.542$ ms, data storage computation incurs 3$T_m$ + 3$T_h$ + (x+1)$T_a \approx 39.444$ ms, data request costs 2$T_h$ + 2$T_m \approx 59.534$ ms, data/key transformation costs 4$T_m$ + (x+1)$T_a$ + 8$T_h \approx 41.254$ ms, and data decryption incurs a cost of 4$T_h \approx 0.016$ ms.

\begin{table}[!t]
	\caption{Computation Cost Comparsion(ms)\label{tab:table1}}
	\centering
	\begin{tabular}{|c|c|c|c|c|}
		\hline
		Schemes/& $SD_i$ &  $ES_i$ &  $ES_j$ &  $SD_j$\\
		\hline
		Zhong \textit{et al.} \cite{liu2024dkgauth} & - & - & 339.852 & 295.263\\
		\hline
		Wang \textit{et al.} \cite{wang2023lightweight} & 84.797 & 83.253 & 36.938 & 113.163\\
		\hline
		Mao \textit{et al.} \cite{mao2023locally} & 90.205 & 85.887 & - & -\\
		\hline
		Hu \textit{et al.} \cite{hu2024security} & 84.793 & 180.157 & 38.884 & -\\
		\hline
		Wang \textit{et al.} \cite{wang2024blockchain} & - & 73.028 & 114.852 & 236.874\\
		\hline
		Ours & 56.542 & 39.444 & 41.254 & 59.55\\
		\hline
	\end{tabular}
\end{table}

\begin{figure}[!t]
	\centering
	\includegraphics[width=0.47\textwidth]{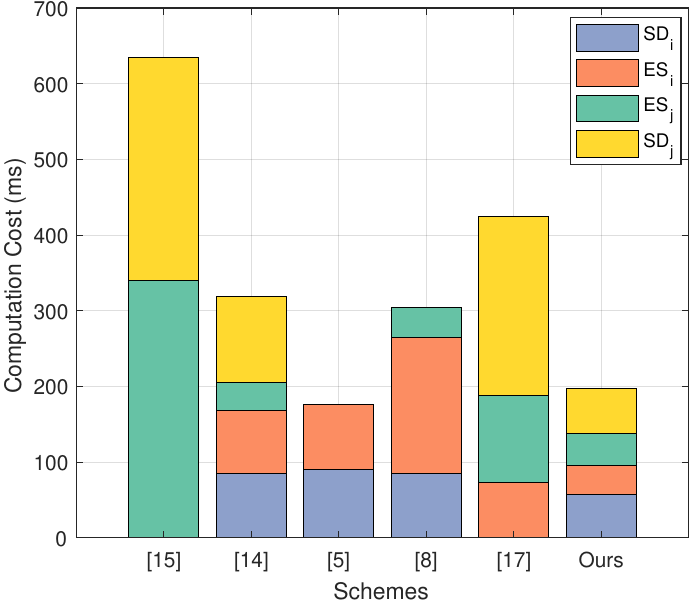}
	\caption{Computation cost comparison.}
	\label{fig_1}
	\vspace{-1.6em}
\end{figure}

The theoretical analysis of computational overhead for various schemes is shown in Table III, the actual computational overhead is presented in Table IV, and the comparison is illustrated in Figure 3.

It is evident that our scheme provides a complete data-sharing process, unlike other schemes that address only specific steps, such as data upload or data sharing. It is worth noting that while the total overhead of our scheme is slightly higher than that of Mao’s scheme, this is because Mao’s scheme\cite{mao2023locally} performs data decryption operations entirely at Information Processing Center (IPC), and the details on how data users obtain the data are not discussed in the paper. Additionally, as shown in Table IV and Figure 3, our scheme is more device-friendly and suitable for resource-constrained devices.

Additionally, our scheme supports batch authentication, which we analyzed and compared with other existing schemes.

Zhong \textit{et al.} \cite{zhong2021broadcast}’s scheme requires 3$T_{bp}$ + 2$nT_m$ + 2$nT_{sm}$ + $(4n-3)T_a$ + 3$nT_h$ + 4$T_{gtmul} \approx 18.689n + 61.278$ ms. Wang \textit{et al.} \cite{wang2023lightweight} report a batch authentication overhead of $nT_{sm}$ + $(n+2)T_m$ + $(2n+2)T_a$ + $2nT_h \approx 9.345n+18.538$ ms. In Mao \textit{et al.} \cite{mao2023locally}, the overhead for batch verification of aggregated signatures is 2$T_{bp}$ + $nT_a$ + $(2n+1)T_m \approx 18.503n+49.924$ ms. Hu \textit{et al.} \cite{hu2024security} present a scheme where the overhead is 2$T_{bp}$ + $(2n+1)T_a$ + $nT_h$ + $nT_m \approx 9.305n+40.725$ ms. Lastly, Wang \textit{et al.} \cite{wang2024blockchain} describe a batch authentication overhead of $(2n+2)T_m$ + $(2n+1)T_a$ + $nT_h$ + $nT_{sm} \approx 18.578n+18.503$ ms.

In contrast, our scheme achieves a batch authentication overhead of $(n+1)T_m$ + $nT_{sm}$ + $(2n+nx+1)T_a$ + $nT_h \approx 13.619n+9.344$ ms.

\begin{figure}[!t]
	\centering
	\includegraphics[width=0.45\textwidth]{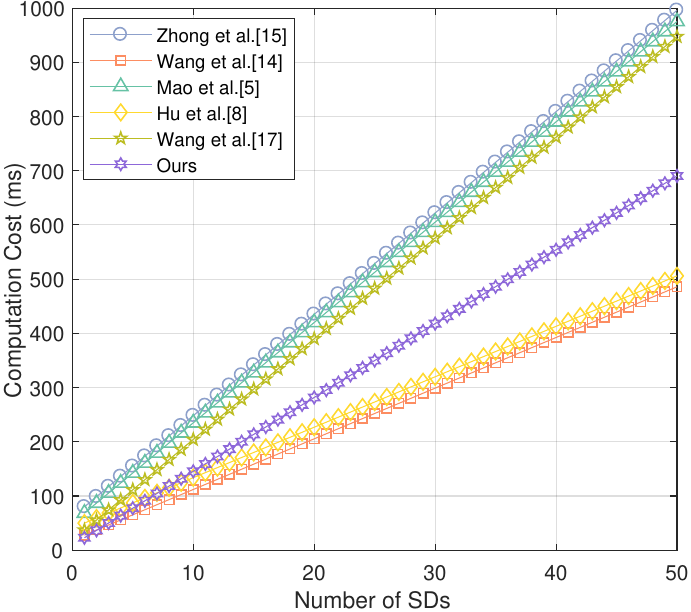}
	\caption{Batch authentication comparison.}
	\label{fig_1}
	\vspace{-1.6em}
\end{figure}

As shown in Figure 4, our scheme has slightly higher batch authentication overhead compared to Wang\cite{wang2023lightweight}'s and Hu\cite{hu2024security}'s schemes. This is because, in Wang's scheme, there is no pseudonym update process during batch authentication; SDs use the same pseudonym for authentication, which can lead to privacy leakage and fails to achieve unlinkability. In contrast, our scheme uses a new pseudonym for each batch authentication, resulting in overhead primarily due to the computation of the public key for the new pseudonym.
In Hu's scheme, the lower computation for signature verification is because part of the computation is offloaded to the SDs. For example, the signature verification for SDs requires 3$T_m+T_h \approx 84.793$ ms, whereas our scheme requires only 2$T_m+4T_h \approx 56.542$ ms. Therefore, our scheme is more suitable for resource-constrained devices.

\subsection{Communication Cost Analysis}

\begin{figure}[!t]
	\centering
	\includegraphics[width=0.46\textwidth]{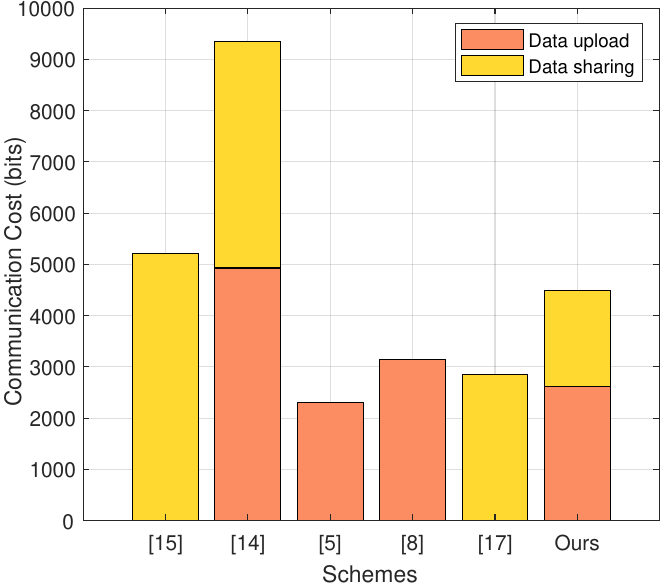}
	\caption{Communication cost comparison.}
	\label{fig_1}
	\vspace{-1.6em}
\end{figure}
As in the experimental setup described above, we set the length of the hash function to 256 bits, the length of the timestamp $T$ to 32 bits, the length of the points on ECC to 256 bits, the length of the device pseudonym to 256 bits, and the length of the index for stored data to 256 bits. For fairness in comparing the schemes, we do not include the length of the uploaded $M$, $M_{type}$, and their ciphertexts when calculating communication overhead. For ease of comparison, we divide the entire process into two parts: data upload and data sharing.

In Zhong \textit{et al.} \cite{zhong2021broadcast}’s scheme, only data download and decryption are considered, without the data upload part. The communication overhead for device signatures and data request is 512 + 512 + 32 $\approx$ 1056 bits, and the communication overhead for TA re-encryption is 256$\times$4 + 512$\times$6 + 32$\times$2 $\approx$ 4160 bits. In Wang \textit{et al.} \cite{wang2023lightweight}’s scheme, both data upload and data download/decryption are included. The communication overhead for the data upload part is 256$\times$7 + 512$\times$6 + 32$\times$2 $\approx$ 4928 bits, and the communication overhead for data download/decryption is 256$\times$5 + 512$\times$6 + 32$\times$2 $\approx$ 4416 bits. In Mao \textit{et al.} \cite{mao2023locally}’s scheme, only data upload is considered, with a communication overhead of 512$\times$3 + 256$\times$3 $\approx$ 2304 bits. In Hu \textit{et al.} \cite{hu2024security}’s scheme, only data upload and on-chain broadcasting are considered, with a communication overhead of 512$\times$4 + 32$\times$2 + 256$\times$4 $\approx$ 3136 bits. In Wang \textit{et al.} \cite{wang2024blockchain}’s scheme, both data re-encryption and sharing processes are included, with a communication overhead of 512 + 256$\times$9 + 32 $\approx$ 2848 bits.

In our scheme, both data upload and data request are considered. The communication overhead for data upload and storage is 512$\times$2 + 256$\times$6 + 32$\times$2 $\approx$ 2624 bits, and the communication overhead for the data request part is 512 + 256$\times$5 + 32$\times$2 $\approx$ 1856 bits.
\begin{table*}[h]
	\caption{Comparison of Security and Functionality Features}
	\centering  
	\begin{threeparttable}
	\begin{tabular}{ccccccc}
		\hline
		Security and Functionality Features &Zhong \textit{et al.}\cite{zhong2021broadcast} &Wang \textit{et al.}\cite{wang2023lightweight} &Mao \textit{et al.}\cite{mao2023locally} &Hu \textit{et al.}\cite{hu2024security} & Wang \textit{et al.}\cite{wang2024blockchain}& Ours\\ 
		\hline
		Confidentiality  & \Checkmark &\Checkmark&\Checkmark &\Checkmark &\Checkmark  & \Checkmark\\
		Key Security  &\Checkmark &\XSolidBrush& \XSolidBrush&\XSolidBrush &\XSolidBrush  & \Checkmark\\
		Data Traceability  &\XSolidBrush &\XSolidBrush& \Checkmark& \Checkmark&\XSolidBrush  & \Checkmark\\
		Perfect Forward and Backward Secrecy  &\Checkmark &\Checkmark&\XSolidBrush &\XSolidBrush & \XSolidBrush & \Checkmark\\
		Conditional Privacy Protection  & \Checkmark&\Checkmark &\XSolidBrush&\Checkmark &\Checkmark  & \Checkmark\\
		Anonymity &\Checkmark &\Checkmark&\XSolidBrush &\Checkmark &\Checkmark  & \Checkmark\\
		Unlinkability &\Checkmark &\XSolidBrush&\XSolidBrush &\XSolidBrush &\Checkmark  & \Checkmark\\
		Resistance to Impersonation Attacks  &\Checkmark &\Checkmark&\Checkmark &\Checkmark & \Checkmark & \Checkmark\\
		Resistance to Replay Attacks  &\XSolidBrush &\Checkmark&\XSolidBrush &\Checkmark &\Checkmark  & \Checkmark\\
		Resistance to Modification Attacks  &\Checkmark &\Checkmark&\Checkmark &\Checkmark &\Checkmark  & \Checkmark\\
		\hline
	\end{tabular}
    \begin{tablenotes}
    	\footnotesize
    	\item[1] \Checkmark indicates that the scheme is secure or provides that functionality, and \XSolidBrush indicates that the scheme is insecure or does not provide that functionality.
    \end{tablenotes}
\end{threeparttable}
\end{table*}
As shown in Figure 5 and based on the above calculations, we can see that the communication overhead of our scheme in the data upload and storage parts is approximately 53.24\% of Wang \textit{et al.} \cite{wang2023lightweight}’s scheme (2624/4928 $\approx$ 53.24\%), 113.89\% of Mao \textit{et al.} \cite{mao2023locally}’s scheme (2624/2304 $\approx$ 113.89\%), and 83.67\% of Hu \textit{et al.} \cite{hu2024security}’s scheme (2624/3136 $\approx$ 83.67\%). It should be noted that in Mao \textit{et al.} \cite{mao2023locally}’s scheme, a single piece of data is transmitted in a point-to-point manner, unlike other schemes where data is stored for sharing among multiple users. This reduces communication overhead, but if Mao’s scheme aims to share data with multiple users, the data producer would have to frequently send a large amount of data, leading to higher communication overhead.

On the other hand, the communication overhead of our scheme in the data request and sharing parts is approximately 35.58\% of Zhong \textit{et al.} \cite{zhong2021broadcast}’s scheme (1856/(1056+4160) $\approx$ 35.58\%), 42.03\% of Wang \textit{et al.} \cite{wang2023lightweight}’s scheme (1856/4416 $\approx$ 42.03\%), and 65.17\% of Wang \textit{et al.} \cite{wang2024blockchain}’s scheme (1856/2848 $\approx$ 65.17\%).

From the above analysis, we can conclude that our scheme has lower communication overhead in both data upload and data sharing compared to the other schemes.

\subsection{Comparison of Security and Functionality Features}
In this subsection, we will compare the security and functionality of our scheme with those of others. The comparison results are shown in Table V. It can be observed that our scheme exhibits higher security performance and offers more functionalities.

\section{Conclusion}

In this paper, we present an efficient, secure blockchain-based data-sharing scheme designed to address the security and efficiency challenges of cross-domain data sharing in IoT environments. By adopting a distributed key generation method, our scheme eliminates single point of failure and enhances the security and reliability of the data-sharing process. The independent pseudonym generation and  key updates improve authentication flexibility and reduce the computational burden on resource-constrained IoT devices. Additionally, the scheme provides a comprehensive data-sharing framework, ensuring secure data uploading, storage, and sharing, while maintaining data traceability, integrity, and privacy. Our security analysis confirms that the proposed scheme is resistant to a wide range of attacks, and performance evaluations demonstrate that it incurs lower computational and communication overhead than existing solutions. This makes it highly suitable for resource-constrained IoT applications that require both efficiency and security.\par 
Looking ahead, future work will focus on flexible and adaptive granularity in data-sharing schemes, allowing data users to dynamically adjust parameters such as the time range based on their specific needs when requesting data. Encryption keys will also become more flexible, allowing seamless key transfer and minimizing key redundancy.

\bibliographystyle{ieeetr}

\bibliography{references}

\begin{IEEEbiography}
	[{\includegraphics[width=0.9in,height=1.25in,clip,keepaspectratio]{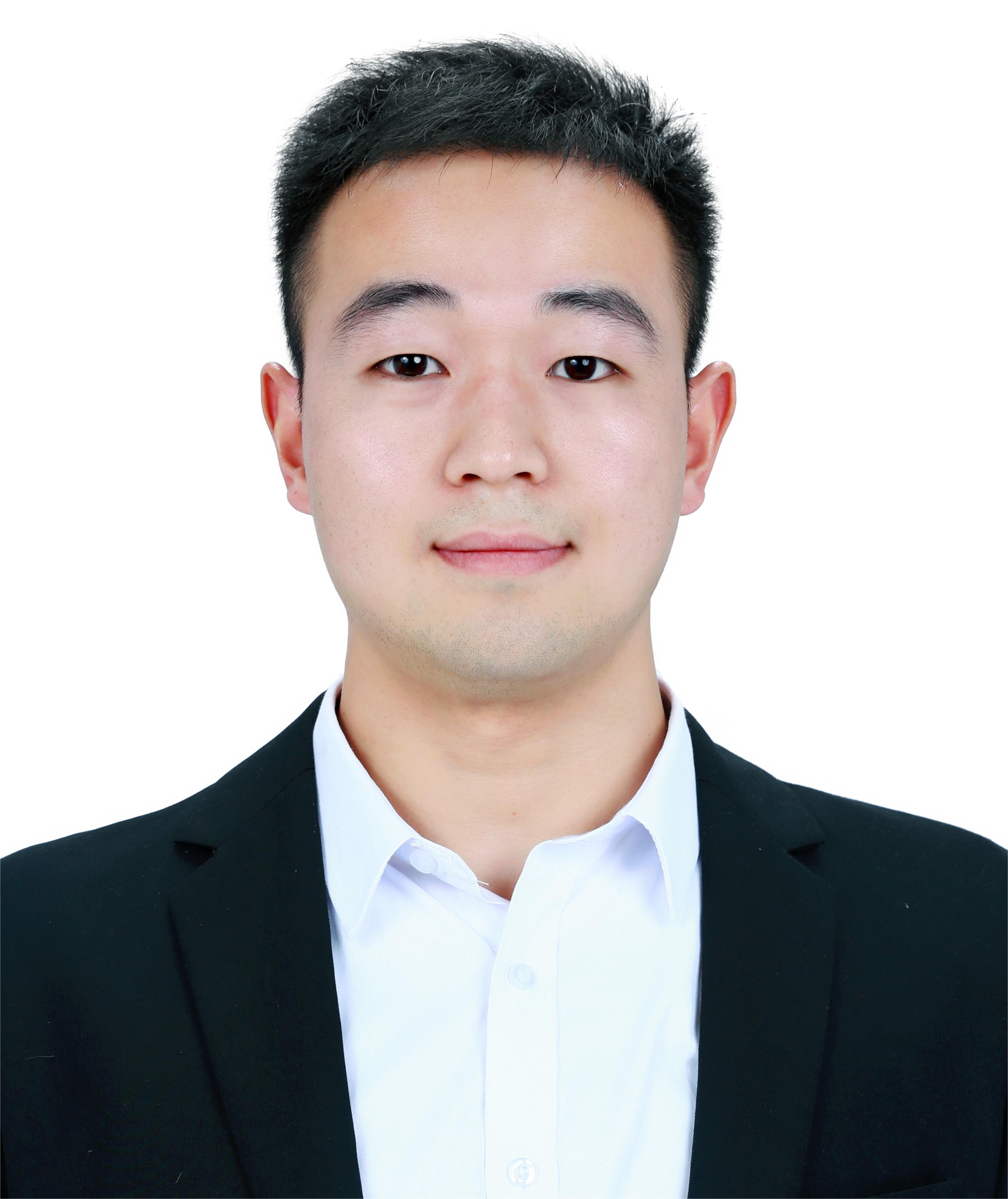}}]
	{\\ \\ \\ Kexian Liu}is currently pursuing the Ph.D. degree with the
	State Key Laboratory of Networking and Switching Technology, BUPT, Beijing, China. His research focuses on future network architecture and network security.

\end{IEEEbiography}

\begin{IEEEbiography}
	[{\includegraphics[width=1in,height=1.25in,clip,keepaspectratio]{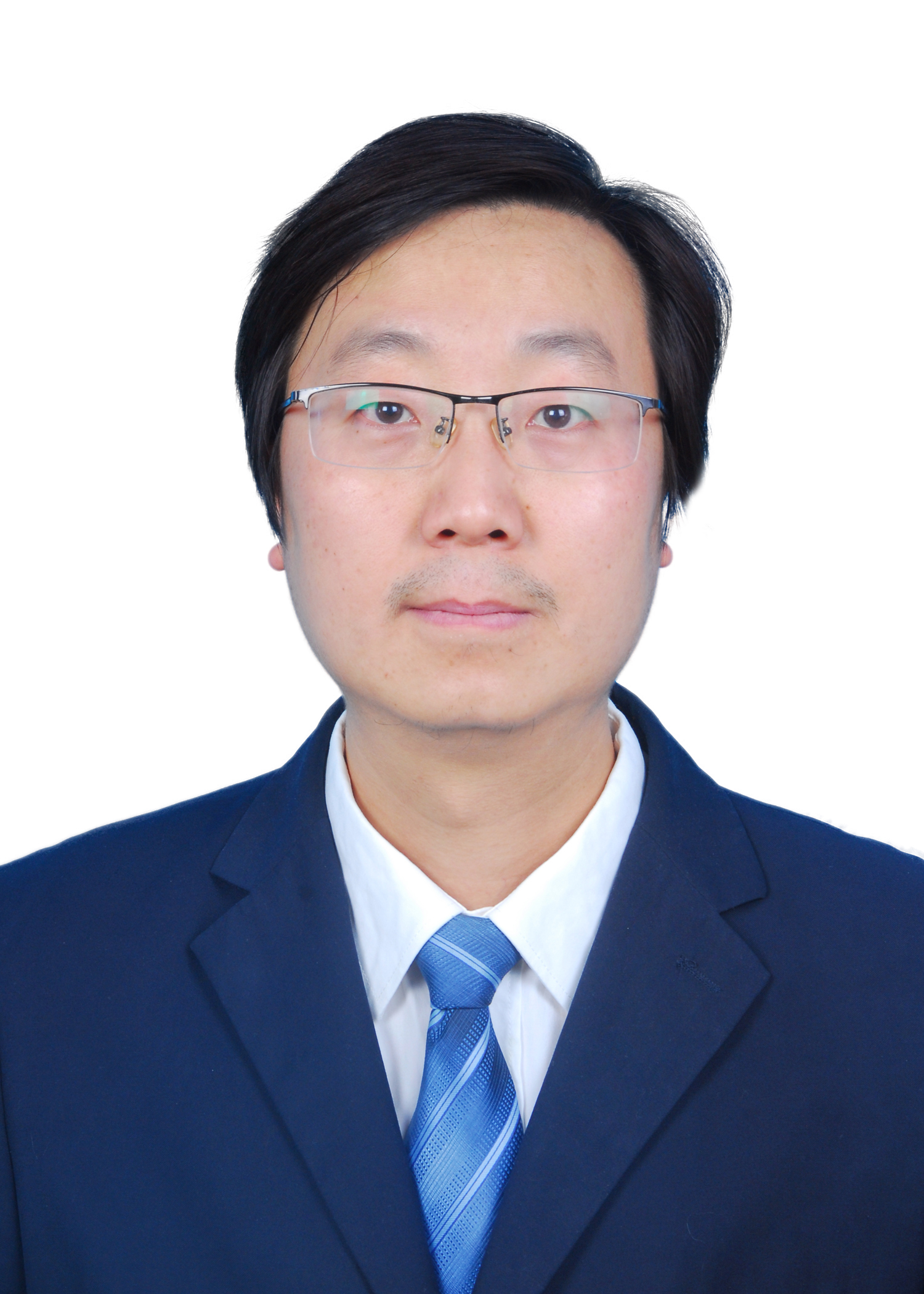}}]
	{Jianfeng Guan}(Member, IEEE) received the B.S. degree in telecommunication engineering from Northeastern University, Shenyang, Liaoning Province, China in 2004, and Ph.D degree in Communication and Information System from Beijing Jiaotong University, Beijing, China in 2010. From 2010 to 2015, he was a lecture with Institute of Network Technology, Beijing University of Posts and Telecommunications. Since 2016, he has been an Assistant Professor. He is the author of more than 100 articles and more than 70 inventions. His research interests include future network architecture, network security and mobile Internet. Dr. Guan is a recipient of several Best Paper Awards from ACM Mobility conference 2008, IC-BNMT2009, and Mobisec2018. He servers as TPC member for WCNC2019, ICC 2018 CCNCP, Globecom 2018 CCNCPS, MobiSec 2016-2019, INFOCOM MobilWorld 2011, 2015-2017, ICCE 2017. He also is a reviewer for Journals such as TVT, TB, CC, CN, JNCA, FGCS, ACCESS, INS, SCN, IJSSC, IJAHUC, JoWUA. 
	%	received the BE degree in telecommunication engineering from Northeastern University, Shenyang, China, in 2004, and the PhD degree in communication and information system from Beijing Jiaotong University, Beijing, China, in 2010. He is currently anassociate professor at the Beijing University of Posts and Telecommunications, China. His research interests include future network architecture, network security, and mobile Internet.

\end{IEEEbiography}

\begin{IEEEbiography}
	[{\includegraphics[width=1in,height=1.25in,clip,keepaspectratio]{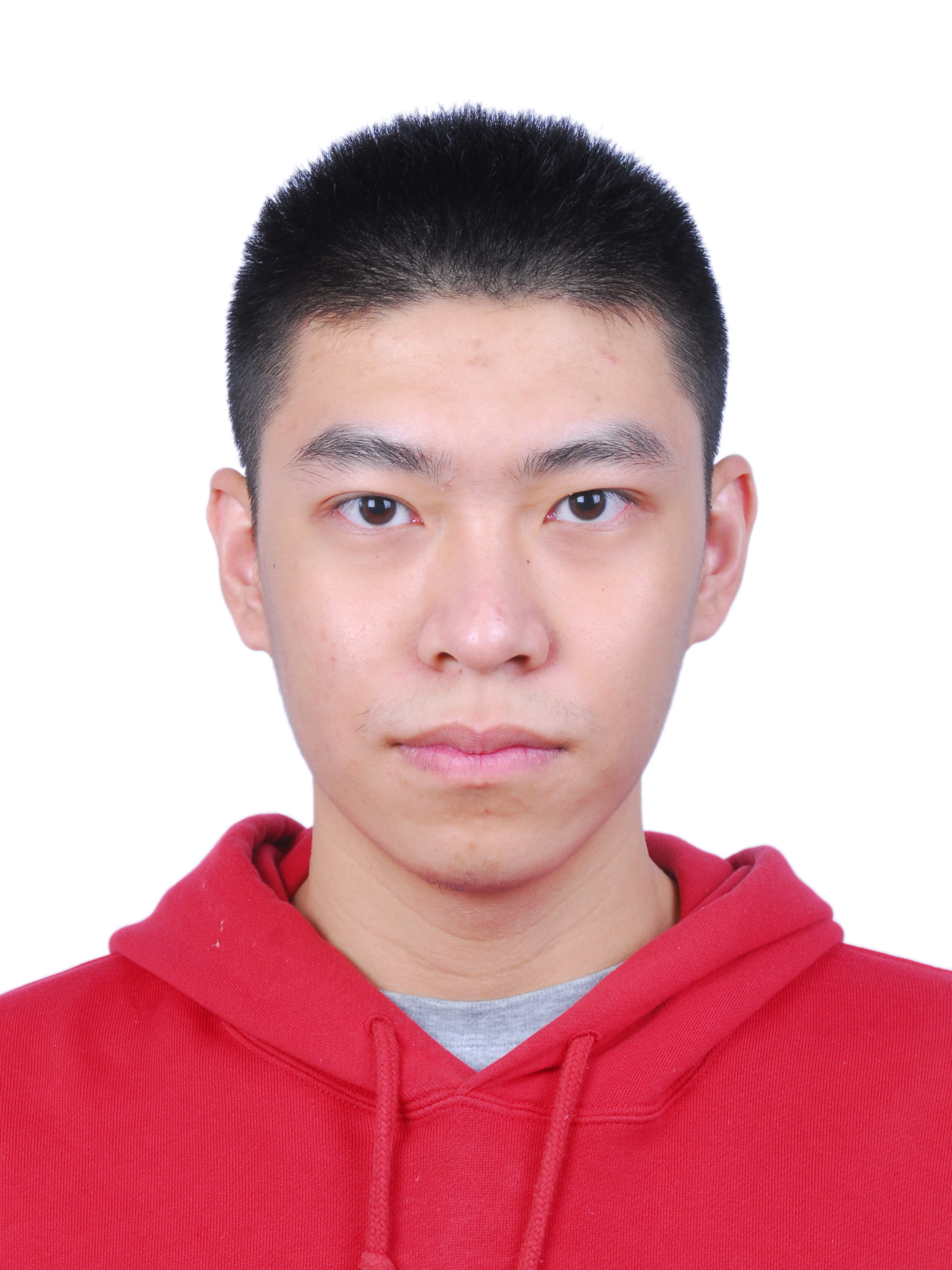}}]
	{\\ \\ \\Xiaolong Hu} is currently pursuing the master degree with the
	State Key Laboratory of Networking and Switching Technology, BUPT, Beijing, China. His research focuses on future network architecture and network security.
	
\end{IEEEbiography}
%\begin{IEEEbiography}
%	[{\includegraphics[width=1in,height=1.25in,clip,keepaspectratio]{zhang1.jpg}}]
%	{\\ \\ \\Jing Zhang} is currently pursuing the Ph.D. degree with the
%	State Key Laboratory of Networking and Switching Technology, BUPT, Beijing, China. Her research focuses on future network architecture and intelligent routing.
%	
%\end{IEEEbiography}

\begin{IEEEbiography}
	[{\includegraphics[width=1in,height=1.25in,clip,keepaspectratio]{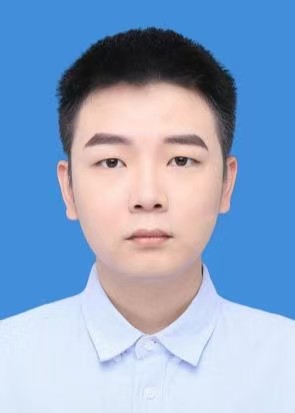}}]
	{\\ \\ \\Jianli Liu} is currently pursuing the master degree with the
	State Key Laboratory of Networking and Switching Technology, BUPT, Beijing, China. His research focuses on authenticaiton.

\end{IEEEbiography}

\begin{IEEEbiography}
	[{\includegraphics[width=1in,height=1.25in,clip,keepaspectratio]{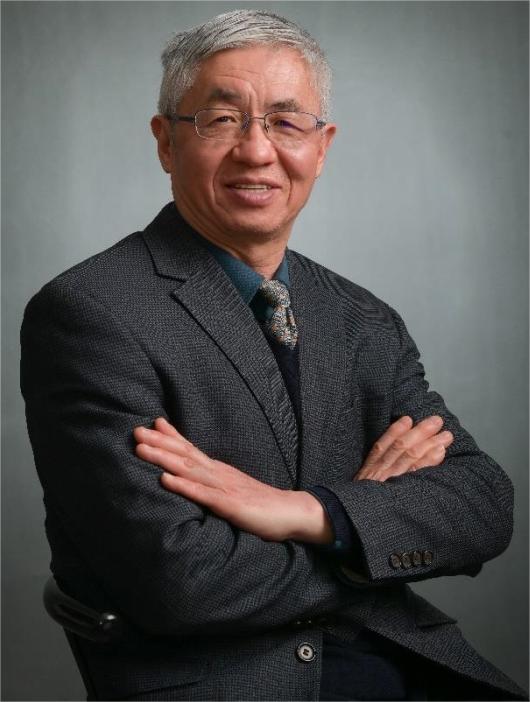}}]
	{Hongke Zhang} (Fellow, IEEE) received the Ph.D. degree in communication and information system from the University of Electronic Science and Technology of China, Chengdu, China, in 1992. He is currently a Professor with the School of Electronic and Information Engineering, Beijing Jiaotong University, Beijing, China, where he currently directs the National Engineering Center of China on Mobile Specialized Network. He is an Academician of China Engineering Academy, Beijing, and the Co-Director of the PCL Research Center of Networks and Communications, Peng Cheng Laboratory, Shenzhen China. His current research interests include architecture and protocol design for the future Internet and specialized networks. Prof. Zhang currently serves as an Associate Editor for the IEEE TRANSACTIONS ON NETWORK AND SERVICE MANAGEMENT and IEEE INTERNET OF THINGS JOURNAL.
	
\end{IEEEbiography}

\end{document}